\documentclass[%
 aip,
 amsmath,amssymb,
 reprint,%
]{revtex4-1}

\usepackage{xcolor}
\usepackage{graphicx}%
\usepackage{dcolumn}%
\usepackage{bm} %
\usepackage{verbatim}

\usepackage[utf8]{inputenc}
\usepackage[T1]{fontenc}
\usepackage{mathptmx}
\usepackage{etoolbox}
\usepackage[textsize=scriptsize, colorinlistoftodos, figwidth=8cm]{todonotes}

\usepackage{marginnote} %

\usepackage[separate-uncertainty]{siunitx}

\usepackage{booktabs}
\usepackage{hyperref}

\newcommand{\tabcref}[1]{\hyperref[#1]{Table~\ref*{#1}}}
\newcommand{\figcref}[1]{\hyperref[#1]{Fig.~\ref*{#1}}}
\newcommand{\eqcref}[1]{\hyperref[#1]{Equation~\ref*{#1}}}
\usepackage[above, below]{placeins}

\usepackage{tikz}
\usepackage{graphicx}
\usepackage{subcaption}
\usetikzlibrary{shapes.geometric, arrows.meta, positioning, calc} 
\usepackage[ruled,vlined,linesnumbered]{algorithm2e} 
\usepackage{geometry}
\usepackage{fontspec}
\usepackage{xeCJK}

\makeatletter
\def\@email#1#2{%
 \endgroup
 \patchcmd{\titleblock@produce}
  {\frontmatter@RRAPformat}
  {\frontmatter@RRAPformat{\produce@RRAP{*#1\href{mailto:#2}{#2}}}\frontmatter@RRAPformat}
  {}{}
}%
\makeatother

\usepackage{tabularx}
\usepackage{booktabs}
\usepackage{makecell}
\usepackage{tabulary}

\begin{document}

\title{\textbf{Bounce or coalescence : a physical learning frame} 
}%

\author{J. H. Xu{(徐伽好)}}
\author{Z. L. Wang{(王志亮)}}%
 \email{wng\_zh@i.shu.edu.cn}
\affiliation{Shanghai Institute of Applied Mathematics and Mechanics, Shanghai Key Laboratory of Mechanics in Energy Engineering, School of Mechanics and Engineering Science, Shanghai University, Yanchang Road 149, Shanghai, 200072, P.R. China
}%

\date{\today}

\begin{abstract}
The numerical simulation of coalescence and bouncing between droplets, or between a droplet and a liquid surface, remains challenging because of the complex topological changes of the interface. Existing approaches usually rely on interface-capturing methods, adaptive mesh refinement（AMR）, thin-gas-film models, or molecular potential models. However, it remains difficult for these methods to simultaneously resolve the high computational cost, stochastic features, and complex phase behavior arising from material-property contrasts, interfacial dynamics, and coupled spatiotemporal multiscale processes. In this study, we develop an interface-contact simulation framework based on physical criteria and machine-learning-assisted classification to describe coalescence and bouncing within a unified formulation. The framework realizes interfacial coalescence and bouncing through the fusion and generation of multiple volume-of-fluid fields. When adjacent interfaces are predicted to coalesce, multiple VOF fields are collapsed into a single VoF field. When approaching interfaces are predicted to bounce, a single VOF field is regenerated into multiple VOF fields, allowing the interfaces to continue evolving independently. With this treatment, the difficulties associated with topological transition, regime-map identification, increasing computational demand, and stochastic behavior during interfacial approach are separated from the interface-tracking procedure. These decisions are instead assigned to a physics-guided machine-learning model with strong adaptability. This strategy avoids the direct resolution of an ultrathin gas film and reduces the dependence on empirical molecular-force parameters. Simulations of droplet--droplet collisions show that the proposed framework can reproduce both coalescence and bouncing over different impact conditions. By further introducing a drainage-time criterion, the framework is extended to the simulation of droplet impact on a liquid surface. For this problem, the numerical results agree well with both previous experimental observations and the present experiments. Moreover, the framework captures the complete sequence of bouncing followed by subsequent coalescence within a single simulation, which has rarely been achieved in previous numerical studies.These results demonstrate that the proposed framework has strong adaptability for interfacial contact problems and provides a unified modeling route for droplet coalescence, bouncing, and subsequent interfacial evolution through system evolution and learning-based physical criteria.
\end{abstract}

\maketitle

\section{Introduction}
\label{Introduction}

Droplet--droplet collisions and droplet impact on liquid interfaces are representative interfacial interaction problems in multiphase fluid dynamics. They occur widely in spray combustion, atomization and transport, inkjet manufacturing, and cloud microphysics. In these systems, the bouncing, coalescence, and separation of droplets at the local scale directly affect the macroscopic droplet-size distribution, mixing efficiency, heat and mass transfer, and overall system stability. Sirignano \cite{Sirignano1999} emphasized the fundamental role of droplet dynamics in spray combustion, while Lefebvre and McDonell \cite{Lefebvre2017} summarized the influence of droplet size, breakup, and collision on atomization quality and transport performance. From a broader perspective, Lohse \cite{Lohse2022} highlighted the central role of droplet interactions and interfacial topology in natural and industrial multiphase flows. More recently, Eggers, Sprittles, and Snoeijer \cite{EggersSprittlesSnoeijer2025} further pointed out that droplet coalescence is itself a fundamental topological transition of free-surface flows, with relevance extending from emulsions, sprays, and climate-related flows to processes such as material sintering.

For binary droplet collisions, experimental studies have established systematic classifications of collision outcomes and regime maps. Ashgriz and Poo \cite{Ashgriz1990} identified typical regimes such as coalescence, reflexive separation, and stretching separation. Jiang et al. \cite{Jiang1992} examined the effects of the collision parameter and Weber number, and Qian and Law \cite{Qian1997} constructed a classical collision map showing that the surrounding gas properties and ambient pressure can significantly shift the boundary between bouncing and coalescence. Subsequent studies quantified the boundaries and morphology of head-on collisions of equal-size droplets \cite{Pan2008}, revealed the dynamic differences in collisions of unequal-size droplets \cite{Tang2012}, examined size effects on regime transitions \cite{Huang2021}, and further clarified stretching separation and mode transitions at high Weber numbers \cite{AlDirawi2021Stretching}. These studies show that the final outcome of a droplet collision is not governed solely by the macroscopic competition between inertia and capillarity. The drainage of the gas film entrapped between the approaching interfaces also plays a decisive role.

Mechanistically, the boundary between bouncing and coalescence is controlled by the relative timing of gas-film drainage, interfacial approach, and liquid-bridge formation. Gopinath and Koch \cite{Gopinath2002} analyzed the role of gas lubrication in promoting bouncing under weak deformation, and Bach et al. \cite{Bach2004} showed that viscous resistance in the gas film and interfacial deformability jointly determine whether bouncing occurs. For strongly deforming head-on collisions, Zhang and Law \cite{ZhangLaw2011} developed a model including droplet deformation, internal viscous dissipation, rarefied gas effects, and van der Waals forces, and explained the non-monotonic coalescence--bouncing--coalescence transition with increasing Weber number. Li \cite{Li2016} showed that a macroscopic model with appropriate physical closure can predict the bouncing--coalescence boundary without resolving all molecular-scale details. Chubynsky et al. \cite{Chubynsky2020} further demonstrated that gas kinetic effects and intermolecular forces become important when the gas-film thickness approaches the molecular mean free path. In a recent review, Sprittles \cite{Sprittles2024} summarized that droplet bouncing is essentially governed by the dynamics of the microscopic gas film between approaching interfaces, where conventional continuum descriptions often become insufficient.

Once the gas film ruptures and a liquid bridge forms, the system rapidly enters the early-stage coalescence regime. Eggers et al. \cite{Eggers1999} showed that this stage is characterized by local singularity and self-similar dynamics. Eggers, Sprittles, and Snoeijer \cite{EggersSprittlesSnoeijer2025} recently reviewed the theoretical, experimental, and numerical understanding of bridge growth, emphasizing that the scaling laws are controlled not only by the competition between viscosity and inertia but also by the initial contact geometry and the surrounding fluid. In particular, spherical droplets, droplet--pool systems, and confined geometries do not necessarily share a universal bridge-growth law. Droplet collision and coalescence should therefore be viewed as a multiscale topological transition involving millimeter-scale global deformation, micron- or submicron-scale gas-film drainage, and even smaller-scale interfacial interactions.

These multiscale features make direct numerical simulation of the bouncing--coalescence transition particularly challenging. The volume-of-fluid method of Hirt and Nichols \cite{Hirt1981} laid the foundation for simulating large-deformation free-surface flows, while the level-set method of Sussman et al. \cite{Sussman1994} provided another important approach for complex interfacial topology. Scardovelli and Zaleski \cite{Scardovelli1999} reviewed interface-capturing methods and noted that when the separation between interfaces approaches the grid scale, numerical errors can induce nonphysical topological connection. Later, Popinet \cite{Popinet2003,Popinet2009} developed the Gerris/Basilisk framework based on geometric VOF and adaptive tree meshes, which greatly improved the accuracy and efficiency of surface-tension-dominated two-phase flow simulations. Nevertheless, even with adaptive refinement, the ultrathin gas film that determines bouncing remains difficult to resolve directly in practical parametric studies. As emphasized by Sprittles \cite{Sprittles2024}, if microscale gas-film physics is absent from the model, the predicted location and timing of coalescence can become strongly grid dependent. Thus, for macroscopic interface solvers, the key issue is not only how to capture large-scale deformation, but also how to incorporate the effect of the unresolved gas film in a physically consistent manner.

Several strategies have been proposed to alleviate this difficulty, including multiple-marker interface capturing, local coalescence suppression, and empirical or semi-empirical collision models. Coyajee and Boersma \cite{Coyajee2009} used a multiple-marker front-capturing method to suppress numerical coalescence. Kwakkel et al. \cite{Kwakkel2013} introduced coalescence and breakup models into a CLSVOF framework and determined whether coalescence should occur by comparing the contact time with the drainage time. Al-Dirawi and Bayly \cite{AlDirawi2019} further improved the prediction of bouncing regimes by using experimentally informed collision boundaries. Despite these efforts, for problems involving both large-scale deformation and microscale drainage control, it remains difficult to regulate interfacial topology in a stable and physically consistent manner without excessively resolving the nanometric gas film.

A closely related problem is the non-coalescent bouncing and delayed coalescence of a droplet impacting a deep liquid pool or a free liquid surface. Jayaratne and Mason \cite{Jayaratne1964} observed early that an air layer between the droplet and the liquid surface can support non-coalescent bouncing. Later studies examined the bouncing time and dynamics of droplets on liquid surfaces \cite{Zou2011}, the probability and critical conditions for bouncing \cite{Zhao2011}, and the transition boundary between bouncing and coalescence on deep liquid pools \cite{Wu2020}. Tran et al. \cite{Tran2013} experimentally revealed the sequence of gas-film formation, rupture, and bubble entrapment during droplet impact on a liquid surface. Sprittles \cite{Sprittles2024} further noted that droplet--pool, droplet--film, and droplet--solid impact problems share the same thin-gas-film-controlled mechanism, in which the competition between gas-film drainage and interfacial approach determines whether bouncing, delayed coalescence, or direct contact occurs. More recently, Alventosa et al. \cite{Alventosa2023} showed that, at low Weber numbers, the coupled deformation of the droplet and pool together with inertial--capillary bouncing dynamics can produce pronounced non-coalescent bouncing. Compared with binary droplet collisions, droplet impact on a liquid pool involves stronger deformation of the free surface, and the interfaces may approach again or even repeatedly after the first impact. Therefore, an instantaneous geometric criterion is generally insufficient to describe the full pathway of first impact, short-time bouncing, re-approach, and final coalescence.

Overall, accurately capturing interfacial bouncing and coalescence remains challenging for three main reasons. First, the topological evolution between approaching interfaces is highly complex and is difficult to resolve accurately within conventional interface-capturing methods~\cite{Scardovelli1999,Popinet2009,Coyajee2009,Kwakkel2013}. Second, the physical boundary between bouncing and coalescence can be affected by numerical dissipation, mesh resolution, and interface reconstruction errors, especially when the gas film between the interfaces becomes thinner than the local grid scale~\cite{Gopinath2002,Bach2004,ZhangLaw2011,Chubynsky2020,Sprittles2024}. Third, the transition between bouncing and coalescence is sensitive to liquid properties, gas properties, ambient pressure, impact conditions, and initial disturbances, and may therefore exhibit apparent scatter near the transition boundary~\cite{Qian1997,Huang2021,Zhao2011,Sprittles2024}. 

Previous studies have addressed these difficulties mainly through three types of strategies. The first strategy is to use highly refined adaptive meshes, sometimes approaching sub-micrometer or even nanometric length scales, to directly resolve the thin gas film. This approach can improve physical fidelity but leads to a rapid increase in computational cost~\cite{Chubynsky2020,Sprittles2024}. The second strategy is to embed gas-film drainage models, rarefied-gas corrections, or van der Waals force models into numerical algorithms. These models introduce important unresolved microscopic physics, but their parameters and applicability usually depend on the specific liquid, gas, and impact conditions~\cite{ZhangLaw2011,Kwakkel2013,Chubynsky2020}. The third strategy is to use empirical or semi-empirical regime maps to determine whether bouncing or coalescence should occur. Such models are computationally efficient, but their predictive capability is often limited when material properties, size ratios, or environmental conditions differ from those used for calibration~\cite{Qian1997,Huang2021,Zhao2011}. Therefore, although existing approaches have provided important physical insights and useful numerical tools, it remains difficult for a macroscopic VOF solver to simultaneously achieve low computational cost, stable topological control, and adaptability to different interfacial collision conditions.

Droplet impact on a liquid pool further illustrates this difficulty. Previous studies have revealed non-coalescent bouncing, gas-film formation and rupture, and the bouncing--coalescence transition during droplet impact on liquid surfaces~\cite{Jayaratne1964,Zhao2011,Tran2013,Alventosa2023}. However, most existing studies focus on the first bouncing event, the gas-film drainage or rupture process, or the transition condition between bouncing and coalescence. A unified numerical description of the complete sequence of first impact, short-time bouncing, re-approach, and final coalescence within the same macroscopic simulation remains rarely demonstrated. This motivates the development of a topological-control strategy that can maintain interface separation during the first near-contact event and trigger coalescence only when an appropriate physical criterion is satisfied during later re-approach.

In this work, we propose a machine-learning-assisted topological-control framework for interfacial collision problems. The key idea is to separate the hydrodynamic evolution of the interfaces from the topological decision of whether the approaching interfaces are allowed to connect. In this framework, the flow field and interface deformation are still solved by the incompressible Navier--Stokes equations and the VOF transport equations, whereas the topological outcome is determined by a criterion layer informed by experimental data, machine-learning classification, or an effective drainage-time model. This treatment avoids relying solely on extremely fine meshes or fixed microscopic parameters, and provides an updatable route for incorporating new experimental or high-fidelity numerical data into the topological decision.

The present framework is built on the multiple-VOF solver in Basilisk. For binary droplet collisions, the determination of bouncing and coalescence is formulated as a supervised classification problem. The Weber number $We$ and the impact parameter $B$ are used as input features to train a binary machine-learning model. The resulting $We$--$B$ prediction map is then used as the input to the topological-decision layer, which determines whether the multiple VOF tracer fields should remain separated or be merged after interfacial approach. For droplet impact on a liquid pool, the comparison between the accumulated near-contact time and the effective drainage time is further introduced to describe delayed coalescence. The purpose of this treatment is not to directly resolve all lubrication details inside the nanometric gas film, but to reconstruct bouncing, coalescence, and their temporal transition at low computational cost with an updatable criterion, while retaining the macroscopic deformation, oscillation, and bouncing dynamics of the droplet and the pool. Therefore, the present work focuses on three aspects: establishing an executable multiple-VOF topological-control algorithm; constructing and validating a machine-learning bouncing/coalescence criterion based on $We$ and $B$; and further examining the ability of the framework to reproduce the complete process of first bouncing, re-approach, and final coalescence during droplet impact on a liquid pool.

\section{Numerical method and algorithmic framework}
\label{sec:numerical_method}

\subsection{Governing equations and interface capturing}
\label{subsec:governing_equations}

In this study, numerical simulations of gas--liquid two-phase flow during droplet collisions in a gaseous ambient are performed using the open-source Basilisk C framework \cite{Popinet2003,Popinet2009,popinet2015}. The fluids are treated as incompressible Newtonian fluids, and the governing equations are written as
\begin{equation}
\nabla \cdot \mathbf{u} = 0,
\label{eq:ns_continuity}
\end{equation}
\begin{equation}
\rho \left(
\frac{\partial \mathbf{u}}{\partial t}
+\mathbf{u}\cdot\nabla \mathbf{u}
\right)
=
-\nabla p
+\nabla\cdot\left[
\mu\left(\nabla\mathbf{u}+\nabla\mathbf{u}^{T}\right)
\right]
+\sigma \kappa \mathbf{n}\delta_s
+\rho \mathbf{g},
\label{eq:ns_momentum}
\end{equation}
where $\mathbf{u}$ is the velocity vector, $p$ is the pressure, $\rho$ and $\mu$ are the local density and dynamic viscosity, respectively, $\sigma$ is the surface tension coefficient, $\kappa$ is the interfacial curvature, $\mathbf{n}$ is the unit normal vector to the interface, $\delta_s$ is the interfacial Dirac distribution, and $\mathbf{g}$ is the gravitational acceleration. The surface-tension term is written in the continuum-surface-force form, so that the interfacial force is represented as an equivalent body force distributed over the cells adjacent to the interface.

When two droplets are described by a single VOF volume-fraction field $c$, the local material properties of the gas and liquid phases are obtained by linear interpolation,
\begin{equation}
\rho = c \rho_l + (1-c)\rho_g,
\label{eq:rho_single_vof}
\end{equation}
\begin{equation}
\mu = c \mu_l + (1-c)\mu_g,
\label{eq:mu_single_vof}
\end{equation}
where $\rho_l$ and $\mu_l$ are the density and dynamic viscosity of the liquid phase, and $\rho_g$ and $\mu_g$ are those of the gas phase. This single-VOF formulation naturally represents connected liquid regions. However, when two independent droplets approach each other at the grid scale, it can lead to nonphysical coalescence induced by numerical discretization.

When a multiple-VOF formulation is adopted, two volume-fraction fields, $c_1$ and $c_2$, are used to separately track the gas--liquid interfaces of the two droplets, following the double-VOF method \cite{Coyajee2009,He2019NonMonotonic}. Here, $c_1,c_2\in[0,1]$. To describe both the gas and liquid phases, the local density and dynamic viscosity are directly reconstructed from the two volume-fraction fields as
\begin{equation}
\rho=(c_1+c_2)\rho_l+(1-c_1-c_2)\rho_g ,
\label{eq:rho_multi_vof}
\end{equation}
\begin{equation}
\mu=(c_1+c_2)\mu_l+(1-c_1-c_2)\mu_g .
\label{eq:mu_multi_vof}
\end{equation}
Here, $\rho_l$ and $\mu_l$ are the density and dynamic viscosity of the liquid phase, while $\rho_g$ and $\mu_g$ are those of the gas phase. Each volume-fraction field $c_i\,(i=1,2)$ satisfies the conservative VOF advection equation
\begin{equation}
\frac{\partial c_i}{\partial t}+\nabla\cdot(c_i\mathbf{u})=0,
\qquad i=1,2 .
\label{eq:vof_advection}
\end{equation}
In this formulation, the two volume-fraction fields retain the interfacial identities of the two droplets, thereby avoiding nonphysical coalescence caused by coarse grids or discretization errors during interfacial approach. Unlike existing double-marker VOF methods that mainly enforce bouncing, the present framework further allows the two VOF fields to merge when the coalescence condition is satisfied according to the machine-learning criterion or the drainage-time criterion.

Basilisk C discretizes and solves the governing equations on adaptive Cartesian grids \cite{Popinet2003,Popinet2009,popinet2015}. The multiphase interface is reconstructed using the piecewise-linear interface calculation (PLIC) method, and the interface normal is estimated with the mixed Youngs--centered (MYC) scheme \cite{Scardovelli1999,Popinet2009}. Surface tension is treated using a balanced-force continuum-surface-force formulation combined with height-function curvature evaluation, which reduces parasitic currents and improves curvature accuracy in capillary-dominated flows \cite{Popinet2009,Brackbill1992}. The incompressible Navier--Stokes equations are discretized by a finite-volume method, and the velocity--pressure coupling is enforced through a projection method; the advective terms in the momentum equation are discretized using a second-order upwind scheme. The time step is constrained by both the CFL condition and the capillary stability criterion. Adaptive mesh refinement dynamically adjusts the local resolution according to the interface position and flow-field gradients, thereby reducing computational cost while maintaining sufficient interfacial resolution \cite{Popinet2003,Popinet2009,popinet2015}. This numerical framework has shown good accuracy and robustness in canonical two-phase flow problems, including static droplet equilibrium, capillary-wave oscillation, and three-dimensional jet breakup \cite{Popinet2009,popinet2015}.

\subsection{Overview of the criterion-driven topological-control framework}
\label{subsec:criterion_topology_framework}

The standard multiple-VOF method assigns independent volume-fraction fields to different droplets and can therefore numerically prevent automatic merging of distinct interfaces during collision. In this sense, the method essentially represents a forced non-coalescence treatment. For droplet collisions that are known to bounce, this treatment effectively avoids numerical coalescence caused by coarse grids or discretization errors. However, real droplet collisions do not always result in bouncing; they may also lead to coalescence, reflexive separation, or stretching separation. Therefore, simply maintaining interfacial independence with a multiple-VOF method is insufficient for describing the full range of droplet collision behaviors. The more essential issue is to actively determine, under appropriate physical conditions, whether a topological connection between approaching interfaces should be allowed.

Based on this consideration, a criterion-driven topological-control framework is proposed, as schematically shown in Fig.~\ref{fig:framework}. The framework consists of three main layers. The first is the hydrodynamic solver layer, in which Basilisk computes the temporal evolution of interfacial deformation, velocity field, and pressure field. The second is the interface-proximity detection layer, which identifies at each time step whether distinct interfaces enter a near-contact region and extracts the local or global parameters required by the subsequent criterion. The third is the topological-decision layer, which determines, according to prescribed physical criteria, whether the multiple VOF tracer fields should remain separated or be merged. The former corresponds to bouncing or non-coalescent evolution, whereas the latter corresponds to coalescence.

The central idea of this framework is to explicitly decouple hydrodynamic evolution from interfacial topological decision-making. The hydrodynamic evolution is governed by the incompressible Navier--Stokes equations and the multiple-VOF advection equations, whereas the topological decision is controlled by physical criteria such as experimental regime maps, machine-learning classification results, or drainage-time estimates. With this treatment, the solver does not need to directly resolve nanometric gas-film drainage or introduce empirical microscopic parameters such as the Hamaker constant. It can nevertheless reconstruct droplet bouncing, coalescence, and their regime transition at the macroscopic scale. This strategy provides a unified and extensible numerical route for different types of interfacial contact problems.

\subsection{VOF Generation and Fusion Model}

In the present framework, the VOF fields are used not only to capture the gas--liquid interface, but also to control the topological states of different liquid regions. Figure~\ref{fig:framework} illustrates the basic idea of this treatment. When the criterion predicts bouncing, the liquid regions are maintained as separate VOF fields. When the criterion predicts coalescence, the independent VOF fields are fused into a new connected liquid region. Therefore, bouncing and coalescence correspond numerically to VOF generation/separation and VOF fusion/reconstruction, respectively.

For the bouncing process, the corresponding VOF fields are kept independent, which can be written as
\begin{equation}
    c \rightarrow \{c_i,c_j\}.
\end{equation}
This operation means that one liquid region is assigned two independent topological labels, or that two existing VOF fields remain separated. In this case, $c_i$ and $c_j$ are advected by the same velocity field, but they are not reconstructed as a single connected interface. This treatment avoids the premature numerical coalescence that may occur in a standard single-VOF method when two interfaces come into contact at the grid scale. It can therefore be used to describe non-coalescing bouncing in binary droplet collisions and droplet impact on a liquid pool.

For the coalescence process, two independent VOF fields are fused into a new VOF field, which can be written as
\begin{equation}
    \{c_i,c_j\} \rightarrow c_{\mathrm{new}},
    \qquad
    c_{\mathrm{new}}=\min(c_i+c_j,1).
\end{equation}
After fusion, the two originally separated liquid regions are reconstructed as a connected interface and subsequently evolve as a single liquid body. This fusion does not mean that the present model directly resolves the rupture of the nanometric gas film or the early growth of the liquid bridge. Instead, it represents an active topological update at the macroscopic computational scale once the physical criterion indicates that coalescence conditions have been satisfied.

Thus, the proposed topological control can be reduced to two basic operations: bouncing corresponds to generating or maintaining two separated VOF fields from one VOF description, whereas coalescence corresponds to fusing two separated VOF fields into one connected VOF field. Although binary droplet collision and droplet impact on a liquid pool use different physical criteria, both problems can be implemented through these two VOF topological conversions.

\begin{figure*}[!htbp]
    \centering
    \begin{minipage}{0.62\textwidth} 
        \centering
        \textbf{(a) Generalized Simulation Workflow} \par\medskip
        
        \begin{tikzpicture}[
            node distance=0.8cm and 0.5cm, 
            startstop/.style={rectangle, rounded corners, minimum width=3cm, minimum height=0.8cm, align=center, draw=black, thick, fill=white, font=\small\bfseries},
            process/.style={rectangle, minimum width=3.5cm, minimum height=1cm, align=center, draw=black, fill=white, font=\small},
            decision/.style={diamond, aspect=2, minimum width=2.8cm, minimum height=1cm, align=center, draw=black, fill=white, font=\small},
            arrow/.style={thick, -Stealth}
        ]

            \node (start) [startstop] {Time Integration\\ $t^n \to t^{n+1}$};
            \node (vof) [process, below=of start] {Advect Multi-VOF Fields\\ $(C_1, C_2, \dots, C_N)$};
            \node (dec1) [decision, below=of vof] {Proximity Check\\ $d_{min} < \delta_{tol}$?};
            
            \node (calc) [process, below=of dec1] {Extract Governing\\ Parameters $\mathbf{P}$};
            
            \node (dec2) [decision, below=of calc] {Evaluate Criterion\\ $\mathcal{R} =\mathcal{F}(\mathbf{P}) \to \text{Regime}$};
            
            
            \node (bounce) [process, below=of dec2, xshift=-3.0cm] {\textbf{Topology Control:}\\ Maintain Distinct VOFs\\ $(C_i \cap C_j = \emptyset)$};
            
            \node (merge) [process, below=of dec2, xshift=3.0cm] {\textbf{Topology Fusion:}\\ Merge VOF Fields\\ $(C_{new} \leftarrow \sum C_i)$};
            
            \node (solve) [process, below=3.5cm of dec2] {Pressure Projection \&\\ Momentum Update};
            
            \node (check_time) [decision, below=0.6cm of solve] {$t < T_{end}$?};
            \node (end) [startstop, below=0.6cm of check_time] {End};

            \draw [arrow] (start) -- (vof);
            \draw [arrow] (vof) -- (dec1);
            \draw [arrow] (dec1) -- node[anchor=west,font=\footnotesize] {Yes} (calc);
            \draw [arrow] (calc) -- (dec2);
            
            \draw [arrow] (dec1.east) -- node[anchor=south,font=\footnotesize] {No} ++(3.5,0) |- (solve.east);

            \draw [arrow] (dec2.west) -| node[anchor=south east,font=\footnotesize] {Bouncing} (bounce.north);
            
            \draw [arrow] (dec2.east) -| node[anchor=south west,font=\footnotesize] {Coalescence} (merge.north);
            
            \draw [arrow] (bounce.south) -- ([xshift=-1cm]solve.north);
            \draw [arrow] (merge.south) -- ([xshift=1cm]solve.north);
            
            \draw [arrow] (solve) -- (check_time);
            
            \draw [arrow] (check_time.west) -- node[anchor=south,font=\footnotesize] {Yes} ++(-4.5,0) |- (start.west);
            
            \draw [arrow] (check_time.south) -- node[anchor=west,font=\footnotesize] {No} (end.north);

        \end{tikzpicture}
    \end{minipage}%
    \hfill
    \begin{minipage}{0.35\textwidth}
        \centering
        \textbf{(b) Topological Control Algorithm} \par\medskip
        
        \begin{algorithm}[H]
            \SetAlgoLined
            \DontPrintSemicolon
            \SetKwInOut{Input}{Input}
            \SetKwInOut{Output}{Output}
            \footnotesize 
            
            \Input{Fields $u^n, p^n, \{C_i\}^n$}
            \Output{Fields at $t^{n+1}$}
            \BlankLine
            \While{$t < T_{end}$}{
                \tcp{Phase 1: Multi-VOF Advection}
                Advect all $C_i$ (PLIC)\;
                Calc min distance $d_{min}$\;
                \BlankLine
                \tcp{Phase 2: Topological Decision}
                \eIf{$d_{min} < \delta_{prox}$}{
                    Extract parameters $\mathbf{P}$ (e.g., $We, B$)\;
                    Evaluate regime $\mathcal{R} = \mathcal{F}(\mathbf{P})$\;
                    \BlankLine
                    \uIf{$\mathcal{R} == \text{Bouncing}$}{
                        \tcp*[h]{Keep Separate} \\
                        \textbf{Maintain Topologies:}\\
                        Keep $C_i, C_j$ as separate tracers\;
                        (No mass transfer)\;
                    }
                    \Else{
                        \tcp*[h]{Merge} \\
                        \textbf{Fuse Topologies:}\\
                        $C_{new} \leftarrow C_i + C_j$\;
                        Reconstruct single interface\;
                    }
                }{
                    Standard VOF steps\;
                }
                \BlankLine
                \tcp{Phase 3: Flow Solver}
                Calc Surface Tension (using all $C$)\;
                Solve Pressure Poisson\;
                Update Velocity $u^{n+1}$\;
                $t \leftarrow t + \Delta t$\;
            }
            \caption{Generalized Framework}
        \end{algorithm}
    \end{minipage}
    
    \caption{Overview of the proposed \textbf{generalized topological control framework}. (a) The workflow abstracts the collision physics into a criterion function $\mathcal{F}(\mathbf{P})$, where $\mathbf{P}$ represents the governing phase-space parameters (e.g., Weber number, impact parameter). (b) The algorithm explicitly manages the topology: 'Bouncing' maintains distinct VOF tracers ($C_i \cap C_j = \emptyset$), while 'Coalescence' triggers a numerical fusion of the scalar fields.}
    \label{fig:framework}
\end{figure*}

\subsection{Modular representation and extensibility of criterion mapping}
\label{subsec:modular_criterion_mapping}

\begin{figure*}[t]
    \centering
    \includegraphics[width=0.90\textwidth,keepaspectratio]{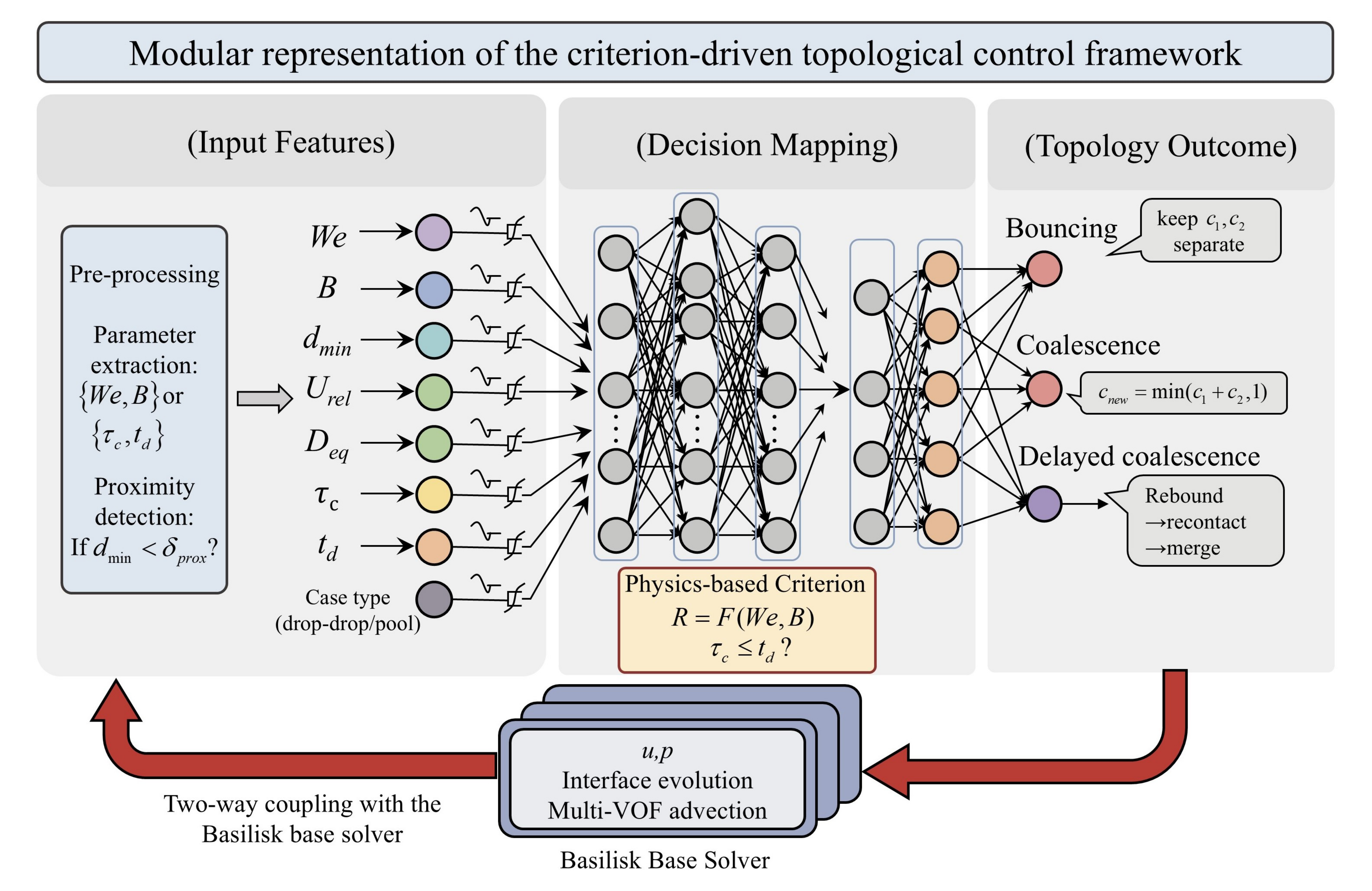}
    \caption{Schematic of the modular representation of the criterion-driven topological-control framework. The input layer contains local geometric, kinematic, and time-scale descriptors, including the Weber number $We$, impact parameter $B$, minimum interfacial distance $d_{\min}$, relative velocity $U_{rel}$, equivalent diameter $D_{eq}$, accumulated near-contact time $\tau_c$, drainage-time threshold $t_d$, and case type. The decision-mapping layer sequentially performs proximity detection, parameter extraction, and physical-criterion evaluation. For droplet--droplet collisions, the program determines bouncing or coalescence according to the regime-map relation $\mathcal{R}=\mathcal{F}(We,B)$. For droplet impact on a liquid pool, the program determines whether to trigger topological connection according to $\tau_c \ge t_d$. The output layer converts the physical outcome into numerical operations, including keeping tracer fields separated, direct merging, and delayed coalescence. The bottom module denotes the two-way coupling between the criterion layer and the baseline Basilisk solver.}
    \label{fig:framework_modular}
\end{figure*}

To clarify the hierarchical structure and unified form of the proposed algorithm, the criterion-driven topological-control method is further organized into a modular representation, as shown in Fig.~\ref{fig:framework_modular}. The purpose of this diagram is to illustrate the correspondence among input information, criterion mapping, and topological output. The baseline flow field is still computed by the Basilisk solver; the machine-learning model does not directly predict the velocity field, pressure field, or interfacial morphology, but acts only on the topological-decision layer after interfacial approach. For binary droplet collisions, a low-dimensional supervised classifier is used to approximate the bouncing/coalescence criterion in the $We$--$B$ plane. For droplet impact on a liquid pool, the physical criterion based on the accumulated near-contact time and the effective drainage time is retained. Therefore, Fig.~\ref{fig:framework_modular} should be interpreted as a machine-learning-assisted topological-decision framework rather than an end-to-end neural-network solver for the flow field.

Figure~\ref{fig:framework_modular} divides the present method into four components: the input layer, the decision-mapping layer, the output layer, and the coupling layer with the baseline solver. The input layer receives the descriptors of the local collision state. These descriptors include geometric, kinematic, and time-scale parameters; therefore, this layer corresponds to local state representation rather than feature learning from flow-field or image data. For droplet--droplet collisions, the program mainly uses parameters such as $We$, $B$, $d_{\min}$, $U_{rel}$, and $D_{eq}$. For droplet impact on a liquid pool, the program focuses on the relation between the near-contact duration $\tau_c$ and the drainage-time threshold $t_d$.

The decision-mapping layer performs proximity detection, parameter extraction, and criterion evaluation. Once two interfaces enter the near-contact region, the program first checks whether the proximity condition is satisfied, then extracts the local state variables corresponding to the current case, and finally calls the appropriate physical criterion. For droplet--droplet collisions, bouncing or coalescence is determined according to the experimentally informed regime-map relation $\mathcal{R}=\mathcal{F}(We,B)$. For droplet impact on a liquid pool, the accumulated near-contact time is used to characterize the duration of the drainage process, and topological connection is triggered when $\tau_c \ge t_d$. This treatment converts the thin-gas-film drainage process, which is difficult to resolve directly, into an executable macroscopic criterion. As a result, the topological evolution of the interface no longer depends solely on local grid resolution and numerical errors.

The output layer converts the physical decision into specific numerical operations. When the criterion predicts bouncing, the program keeps different tracer fields independent. When the criterion predicts direct coalescence, the program merges the tracer fields and updates the interfacial topology. When the criterion predicts delayed coalescence, the program allows the droplet to bounce after the first near contact and then merge after a subsequent re-approach once the drainage condition is satisfied. In this way, a direct correspondence is established between the physical outcome and the numerical update.

The bottom module represents the two-way coupling between the criterion layer and the baseline Basilisk solver. The baseline solver provides the velocity field, pressure field, and multiple-VOF interfacial evolution, whereas the criterion layer returns topological-control instructions to the solver according to the local state. This process forms a closed loop of flow-field evolution, local decision-making, and topological update. This closed-loop structure is one of the key features of the present method. Conventional interface-capturing simulations usually rely on extremely fine meshes to directly resolve thin-film dynamics, whereas the present method reformulates microscopic drainage and contact as an explicit criterion-driven topological-control process, thereby improving model controllability and implementation efficiency.

The modular representation shows that droplet--droplet collisions and droplet impact on a liquid pool involve different physical criteria but share the same algorithmic skeleton: local-state input, proximity identification, criterion evaluation, and topological output. This unified structure gives the present method good extensibility. Future studies can incorporate additional dimensionless parameters, empirical boundaries, or experimental databases into the same decision layer, thereby extending the framework to more complex interfacial contact problems.

\subsection{Interface-proximity detection and local-parameter extraction}
\label{subsec:proximity_detection}

In the present implementation, the minimum interfacial distance $d_{\min}$ is not introduced as an additional physical control parameter, but is used only as a proximity-detection quantity. Its role is to activate the topological-decision module before two interfaces described by different VOF tracer fields come into grid-scale contact. Specifically, at each time step, the program first identifies interfacial cells satisfying $0<c_i<1$, and then estimates the minimum distance $d_{\min}$ between interfaces associated with different tracer fields according to the geometrically reconstructed interface positions. When
\begin{equation}
d_{\min}<\delta_{\mathrm{prox}},
\label{eq:proximity_condition}
\end{equation}
the two interfaces are considered to have entered the near-contact region.

The proximity threshold $\delta_{\mathrm{prox}}$ is chosen to be of the same order as the minimum grid size near the interface,
\begin{equation}
\delta_{\mathrm{prox}}=\alpha \Delta_{\min},
\label{eq:proximity_threshold}
\end{equation}
where $\Delta_{\min}$ is the minimum mesh size near the interface and $\alpha$ is an $O(1)$ constant. Unless otherwise specified, $\alpha=1$ is used in the present simulations. This setting means that the topological criterion is evaluated only when the gas-film thickness between the two interfaces has fallen below the currently resolvable grid scale. Therefore, $\delta_{\mathrm{prox}}$ controls only the activation time of the criterion module and does not directly determine whether the droplets bounce or coalesce. The final topological outcome is still determined by the corresponding physical criterion: for binary droplet collisions, it is determined by the $We$--$B$ regime-map criterion; for droplet impact on a liquid pool, it is determined by the comparison between the accumulated near-contact time and the effective drainage time.

Once the interfaces enter the near-contact region, the solver further extracts the local kinematic parameters. For droplet--droplet collisions, the main parameters include the relative Weber number $We_{\mathrm{rel}}$ and the impact parameter $B$, where
\begin{equation}
We_{\mathrm{rel}}=\frac{\rho_l U_{\mathrm{rel}}^{2} D_{\mathrm{eq}}}{\sigma}.
\label{eq:we_rel}
\end{equation}
Here, $U_{\mathrm{rel}}$ is the relative normal velocity in the near-contact region, and $D_{\mathrm{eq}}$ is the equivalent diameter. The impact parameter $B$ is defined by the offset between the two droplet centroids in the direction perpendicular to their relative motion. For droplet impact on a liquid pool, the pool can be regarded as a large deformable free surface. In the axisymmetric or nearly axisymmetric cases considered here, the impact eccentricity can be neglected. Therefore, $B$ is no longer used as a primary control parameter; instead, the triggering of coalescence is controlled by comparing the accumulated near-contact time with the drainage time.

It should be emphasized that the interface-proximity detection layer is responsible only for identifying when the interfaces enter the applicable range of the criterion; it does not directly determine the collision outcome. The actual topological result is still prescribed by the subsequent physical criterion. This two-layer structure separates proximity identification from physical decision-making, which makes the algorithmic logic clearer and facilitates future extensions to unequal-size droplets, off-center collisions, wall impacts, and other more complex interfacial contact problems.

\subsection{Droplet--droplet collision criterion based on a machine-learning regime map}
\label{subsec:ml_regime_map_criterion}

For binary droplet collisions, a machine-learning regime map in the $We$--$B$ plane is used to determine the topological decision after interfacial near contact. Here, $We$ denotes the Weber number defined based on the relative impact velocity, and $B$ denotes the impact parameter. These two quantities are commonly used as classification parameters in experimental regime maps of binary droplet collisions~\cite{Qian1997,Huang2021}. In the present framework, $We$ and $B$ are used as the only input variables for the droplet--droplet collision criterion, consistent with the input layer shown in Fig.~\ref{fig:framework_modular}. The classifier output contains only two labels: bouncing or coalescence. These two labels correspond directly to two topological operations in the multiple-VOF solver, namely keeping the tracer fields separated or triggering tracer-field merging.

The training data are taken from bouncing and coalescence cases reported by Qian and Law~\cite{Qian1997} and Huang and Pan~\cite{Huang2021}. The real experimental training set contains $66$ bouncing samples and $26$ coalescence samples. Since the experimental data are sparse in some regions of the $We$--$B$ plane, additional boundary-extended samples are generated according to the reported experimental transition boundary. The extended dataset contains $88$ bouncing samples and $103$ coalescence samples, and is used only to regularize the decision boundary in sparsely sampled regions. During training, the extended samples are assigned lower weights than the real experimental samples and are not used for model testing. An independent simulation test set, consisting of $60$ bouncing cases and $45$ coalescence cases, is used only to evaluate the classification capability of the trained criterion.

Because the input space is two-dimensional and the available dataset is limited, a radial-basis-function support vector classifier (RBF-SVC) is adopted to construct the decision boundary. This model is suitable for generating a smooth nonlinear classification boundary in a low-dimensional feature space, and avoids introducing an unnecessary deep neural network for the present tabular data. Before training, the two input variables are normalized to avoid the influence of dimensional and numerical-scale differences on the classification result. After training, the classifier provides a nonlinear boundary between the bouncing and coalescence regions in the $We$--$B$ plane.

The resulting machine-learning regime map is not used merely as an offline post-processing tool, but is embedded into the topological-decision layer of the multiple-VOF framework. Once two droplet interfaces enter the near-contact region, the program evaluates the corresponding $We$ and $B$ for the current condition and queries the trained classifier. If the predicted outcome is bouncing, the two VOF tracer fields remain separated and continue to evolve independently. If the predicted outcome is coalescence, the program performs the tracer-field merging operation.
After this operation, the merged liquid region evolves as a single connected droplet under the Navier--Stokes equations and the VOF transport equation.

It should be clarified that the main purpose of constructing the machine-learning regime map in this study is to examine whether a data-driven bouncing/coalescence criterion can be effectively embedded into the multiple-VOF topological-control framework and further converted into executable interfacial topological operations. Therefore, the present binary classifier using $We$ and $B$ as input variables is not used merely as an offline regime map, but serves as the topological-decision layer for validating the feasibility of the proposed algorithm. The choice of $We$ and $B$ is motivated by the fact that they are the most commonly used control parameters in experimental regime maps of binary droplet collisions. They can be directly connected to available experimental data and are suitable for testing whether the multiple-VOF solver can switch between the two representative topological outcomes of bouncing and coalescence. In this sense, the machine-learning regime map converts the physical criterion contained in experimental regime data into a topological-control instruction for the numerical solver, thereby demonstrating the capability of the proposed multiple-VOF framework to drive interfacial separation and merging through an external criterion. With additional experimental or high-fidelity numerical data, the same decision layer can be further extended to a higher-dimensional input space involving variables such as $Oh$, size ratio, gas pressure, and liquid viscosity.

\subsection{Droplet--pool impact criterion based on an effective drainage time}
\label{subsec:effective_drainage_time}

The topological transition during droplet impact on a liquid pool differs significantly from that in binary droplet collisions. At low Weber numbers, an ultrathin gas film is usually entrapped between the impacting droplet and the free surface of the pool. During the first near-contact stage, this gas film can prevent immediate liquid-bridge formation, leading to non-coalescent bouncing. When the interfaces approach again during the subsequent evolution, and the gas film has drained sufficiently, the system may transition from bouncing to delayed coalescence. Previous studies have shown that whether the interfaces eventually coalesce is essentially determined by whether the duration of near contact is long enough for the entrapped gas film to drain to a critical thickness \cite{ZhangLaw2011,Kwakkel2013}.

It should be noted that the drainage model developed by Zhang and Law \cite{ZhangLaw2011} was formulated for head-on collisions of equal-size droplets in a gaseous medium. The model involves large droplet deformation, internal viscous dissipation, rarefied gas effects, and thin-film rupture induced by van der Waals forces. Kwakkel et al. \cite{Kwakkel2013} further introduced the comparison between contact time and drainage time into a multiple-marker CLSVOF framework as a criterion for distinguishing numerical coalescence from physically allowed coalescence. The present study follows the basic idea of these works. However, because droplet impact on a liquid pool differs from binary droplet collision in geometry, free-surface response, and re-approach pathway, the full drainage dynamics of the nanometric gas film are not directly solved. Instead, an effective drainage-time criterion is constructed for the droplet--pool problem.

Once the system enters the near-contact region, the accumulated near-contact time is defined as
\begin{equation}
\tau_c(t)=\int_{t_1}^{t} I_{\mathrm{near}}(t')\,\mathrm{d}t',
\label{eq:tau_c}
\end{equation}
where $t_1$ is the first time at which the proximity condition is satisfied, and $I_{\mathrm{near}}$ is the near-contact indicator function,
\begin{equation}
I_{\mathrm{near}}(t)=
\begin{cases}
1, & d_{\min}(t)<\delta_{\mathrm{prox}},\\
0, & d_{\min}(t)\ge \delta_{\mathrm{prox}}.
\end{cases}
\label{eq:I_near}
\end{equation}
The accumulated near-contact time is used instead of a single continuous contact time in order to describe the multistage pathway of first near contact, bouncing, re-approach, and final coalescence during droplet impact on a liquid pool. With this definition, the cumulative effect of multiple near-contact events on gas-film drainage can be incorporated into the topological criterion.

For the drainage-time threshold, the following effective closure is adopted:
\begin{equation}
t_d=C_d\sqrt{\frac{\rho_l R^3}{\sigma}},
\label{eq:td_eff}
\end{equation}
where $R$ is the radius of the impacting droplet, $\rho_l$ is the liquid density, $\sigma$ is the surface tension coefficient, and $C_d$ is a calibration constant. The time scale used in this expression is the inertial--capillary time scale. Therefore, $t_d$ does not represent a direct resolution of the full microscopic film-drainage process. Instead, it is a lowest-order effective drainage-time model constructed on the basis of the inertial--capillary time. The coefficient $C_d$ absorbs the combined effects of gas-phase damping, local interfacial deformation, unresolved film dynamics, and unavoidable experimental initial disturbances. Its value therefore needs to be calibrated using experimental observations or high-fidelity numerical results.

It should be emphasized that $C_d$ is not a universal constant directly derived from lubrication theory. In the present framework, it is an effective topological waiting-time coefficient. It represents the combined influence of unresolved gas-film drainage, local interfacial deformation, gas-phase damping, and experimental disturbances. Therefore, the present droplet--pool model should be understood as a semi-empirical closure rather than a fully predictive model requiring no calibration. In this study, the corresponding values of $C_d$ are determined from two experimental cases involving water and silicone-oil droplets. The calibrated criterion is then examined to assess whether it can reproduce the sequence of first bouncing, re-approach, and final coalescence within a single numerical simulation.

The topological triggering condition for droplet impact on a liquid pool is therefore written as
\begin{equation}
\tau_c \ge t_d .
\label{eq:coalescence_trigger_pool}
\end{equation}
Accordingly, the topological-control rule for droplet--pool interaction is expressed as
\begin{equation}
\begin{cases}
\tau_c < t_d, & \text{maintain distinct VOFs},\\
\tau_c \ge t_d, & \text{merge VOF fields}.
\end{cases}
\label{eq:pool_topology_rule}
\end{equation}

The physical meaning of the above criterion can be summarized as follows. The condition $d_{\min}<\delta_{\mathrm{prox}}$ ensures that the system has entered the film-dominated regime. When $\tau_c<t_d$, the entrapped gas film is considered not yet drained to the critical state required for liquid-bridge formation, and the droplet can still bouncing or separate temporarily. When $\tau_c\ge t_d$, the gas film is considered to have drained sufficiently for coalescence to be allowed at the macroscopic level, and topological connection is then triggered. With this treatment, the present method can reconstruct bouncing, re-approach, and delayed coalescence during droplet impact on a liquid pool at the macroscopic scale without directly resolving all lubrication details inside the nanometric gas film.

\section{Basic validation of the algorithm: binary droplet collision}
\label{sec:binary_droplet_validation}

\subsection{Grid-independence analysis}
\label{subsec:grid_independence}

\begin{figure}[htbp]
    \centering
    \includegraphics[width=8.5cm]{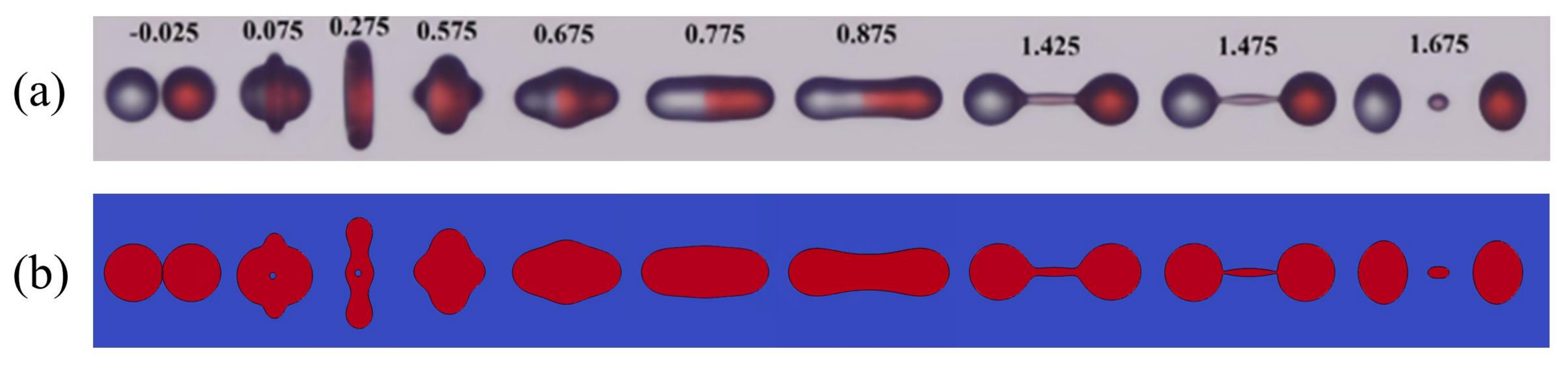}
    \caption{Time sequence of head-on reflexive separation of tetradecane droplets at atmospheric pressure.
    (a) Experimental results of Huang and Pan \cite{Huang2019PRL};
    (b) present numerical results. The parameters are $We=45.92$ and $Oh=0.0376$.}
    \label{fig:binary_morphology}
\end{figure}

To validate the reliability of the present numerical method and topological-control algorithm, we first simulate the head-on reflexive separation experiment of two equal-size droplets reported by Huang and Pan \cite{Huang2019PRL}. Two tetradecane droplets with the same radius, $R_1=R_2=150~\mu{\rm m}$, collide head-on with equal initial velocity magnitudes $U_1=U_2=1.145~{\rm m/s}$ in opposite directions. The corresponding Weber and Ohnesorge numbers are $We=45.92$ and $Oh=0.0376$, respectively. The material properties of the liquid and gas phases are listed in Table~\ref{tab:properties_binary}. Figure~\ref{fig:binary_morphology} compares the experimental images with the present numerical results. The simulation reproduces the main stages observed in the experiment, including droplet compression, radial expansion, liquid-bridge stretching, and final reflexive separation.

\begin{table}[htbp]
\caption{Material properties used in the binary droplet collision validation case.}
\label{tab:properties_binary}
\centering
\begin{tabular}{cccc}
\toprule
 & \makecell{Density \\ (kg/m$^3$)} & \makecell{Viscosity \\ (Pa$\cdot$s)} & \makecell{Surface tension \\ (N/m)} \\
\midrule 
Tetradecane & 759 & $2.05 \times 10^{-3}$ & 0.0260 \\
Gas phase   & 1.2 & $1.78 \times 10^{-5}$ & -- \\
\bottomrule
\end{tabular}
\end{table}

The mesh size is characterized by the maximum refinement level $L$. For the present computational-domain setup, the minimum mesh size corresponding to $L$ is
\begin{equation}
\Delta=\frac{20R_2}{2^L}.
\label{eq:grid_size}
\end{equation}
Thus, the minimum mesh size near the interface decreases as $L$ increases. The adaptive mesh refinement criterion implemented in Basilisk is based on the volume-fraction fields $c_i$ and the velocity field $\mathbf{u}$, so that the interfacial region and regions with large velocity gradients are dynamically refined. To examine the influence of mesh resolution on the numerical results, three maximum refinement levels are considered, corresponding to minimum mesh sizes of $\Delta_{\min}=1.465~\mu{\rm m}$, $0.732~\mu{\rm m}$, and $0.366~\mu{\rm m}$.

\begin{figure}[htbp]
    \centering
    \includegraphics[width=8.5cm]{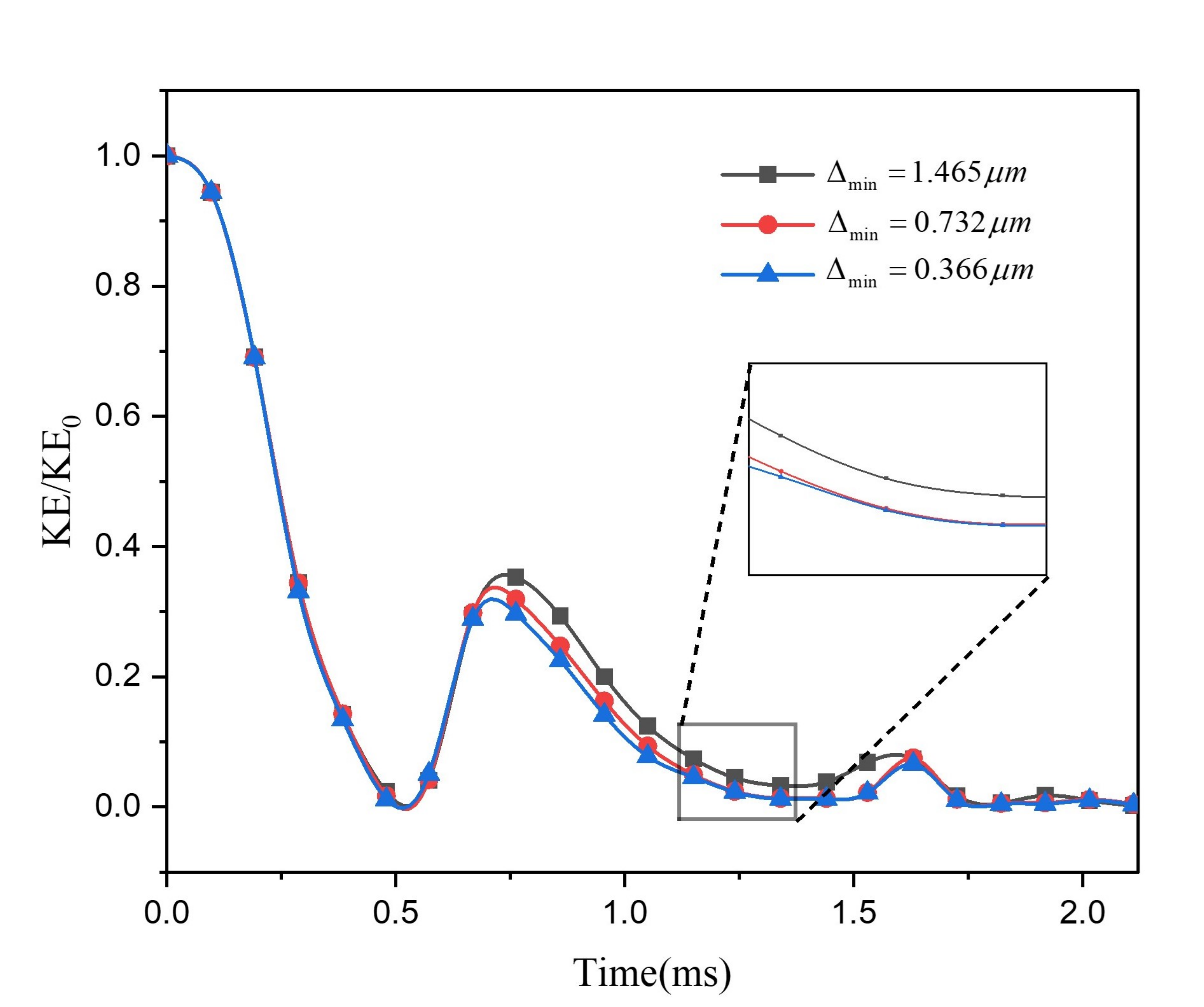}
    \caption{Temporal evolution of the normalized liquid kinetic energy $KE/KE_0$ at different maximum mesh refinement levels.
    The parameters are $We=45.92$ and $Oh=0.0376$.}
    \label{fig:grid_independence}
\end{figure}

Grid independence is evaluated using the temporal evolution of the total kinetic energy of the liquid phase. The liquid kinetic energy is defined as
\begin{equation}
KE=\int_{\Omega_l}\frac{1}{2}\rho_l |\mathbf{u}|^2\,{\rm d}V,
\label{eq:kinetic_energy}
\end{equation}
where $\Omega_l$ denotes the liquid region and ${\rm d}V$ is the cell volume. The total surface energy is evaluated as
\begin{equation}
SE=\sigma A_s,
\label{eq:surface_energy}
\end{equation}
where $A_s$ is the total interfacial area of the liquid droplets, obtained by summing the reconstructed interface areas over all two-phase cells. To quantitatively compare the effect of mesh resolution, Fig.~\ref{fig:grid_independence} shows the temporal evolution of the normalized liquid kinetic energy $KE/KE_0$ at different maximum refinement levels, where $KE_0$ is the initial kinetic energy of the liquid phase. The curves converge as the mesh is refined. Considering both numerical accuracy and computational cost, all subsequent binary droplet collision simulations in this study use the maximum refinement level $L=12$, corresponding to $\Delta_{\min}=0.732~\mu{\rm m}$.

\subsection{Collision morphology and energy evolution}
\label{subsec:morphology_energy_evolution}

\begin{figure}[!htbp]
    \centering
    \includegraphics[width=8.5cm]{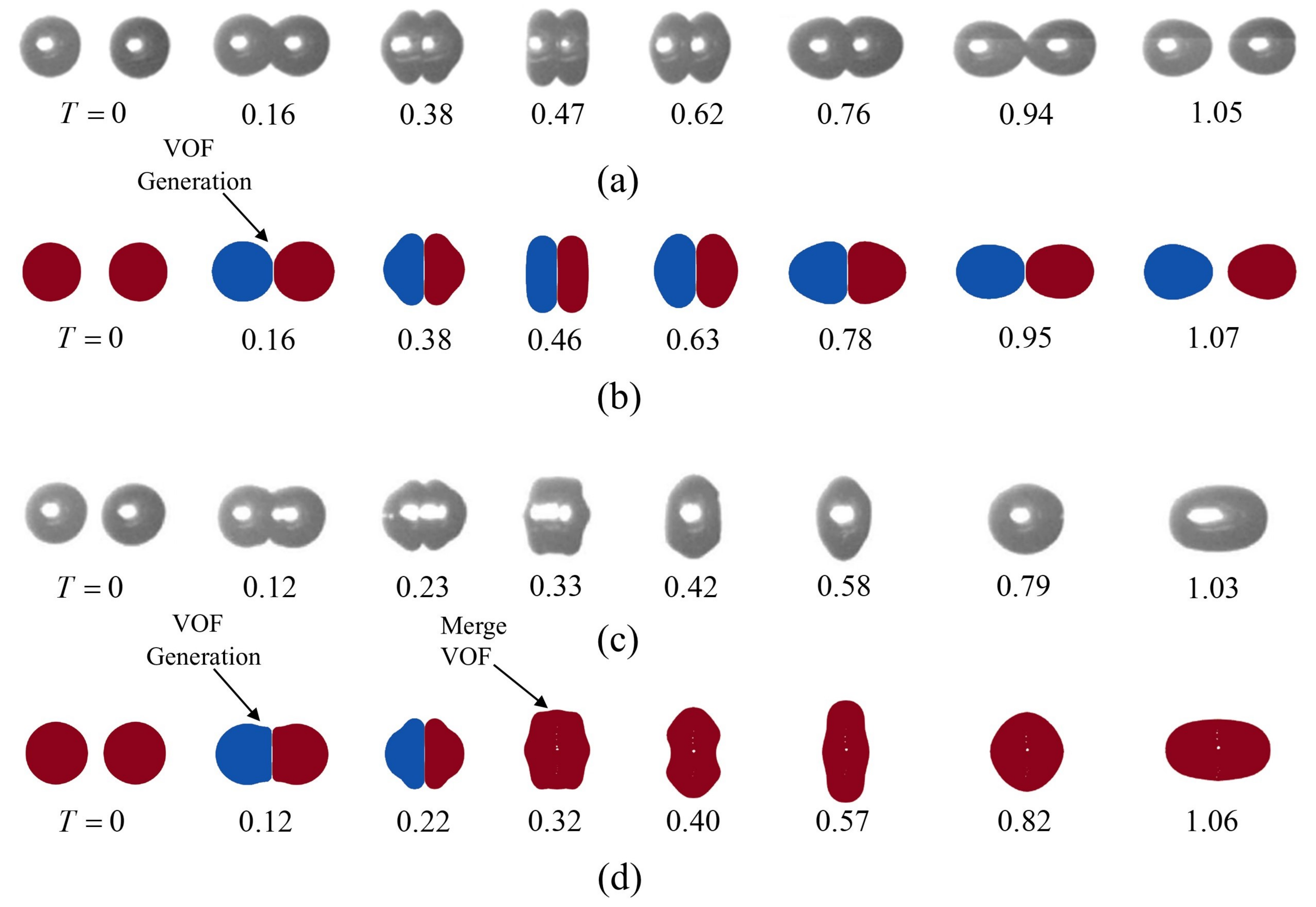}
    \caption{Comparison between experimental and present numerical results for bouncing and coalescence during head-on binary droplet collisions ($B=0$).
    (a) Experimental sequence for the bouncing case at $We=9.33$ and $Oh=0.028$;
    (b) present numerical results corresponding to (a);
    (c) experimental sequence for the coalescence case at $We=13.63$ and $Oh=0.028$;
    (d) present numerical results corresponding to (c).
    The dimensionless time is defined as $T=t/t_c$, where the capillary time scale is $t_c=\sqrt{\rho_l D^3/\sigma}$ with $D=2R_1$. The experimental images are taken from Pan et al. \cite{Pan2008}.}
    \label{fig:pan_morphology}
\end{figure}

To further validate the applicability of the proposed criterion-based topological-control algorithm to the mode transition between bouncing and coalescence, we simulated the head-on collision experiments of two equal-size droplets $(B=0)$ reported by Pan et al. \cite{Pan2008}. Figure~\ref{fig:pan_morphology} compares the collision sequences at the same viscous scale, $Oh=0.028$, but at different Weber numbers.

For $We=9.33$, Figs.~\ref{fig:pan_morphology}(a) and \ref{fig:pan_morphology}(b) show the experimental observations and the numerical results, respectively. Under this low-Weber-number condition, the entrapped gas film does not drain sufficiently after interfacial approach, so no topological connection is triggered and the droplets bounce. As the collision proceeds, the droplets reach their maximum radial spread over $T=0.38$--$0.47$, during which most of the initial kinetic energy is converted into surface energy. The droplets then retract under the action of surface tension and finally separate. The numerical results reproduce well the timing of maximum spreading, recoil, and final detachment.

When the Weber number is increased to $We=13.63$, the system crosses the bouncing--coalescence boundary in the classical experimental regime map of droplet collisions. Figs.~\ref{fig:pan_morphology}(c) and \ref{fig:pan_morphology}(d) present the experimental and numerical results, respectively. Under this condition, the current $We$--$B$ combination satisfies the coalescence criterion, and the algorithm actively triggers the merging of the VOF tracer fields when the two interfaces approach closely. A liquid bridge forms between the droplets at approximately $T\approx0.22$, after which the system enters the large-deformation oscillatory stage of the merged droplet. The numerical results reproduce well the overall morphological evolution after bridge formation and the nonlinear oscillation of the merged droplet.

Overall, the numerical results are in good agreement with the experimental observations during the early collision stage (approximately $T<0.4$). During the later recoil and oscillation stages, small temporal deviations may appear because of numerical dissipation associated with VOF interface reconstruction and the approximation introduced by topological switching. Nevertheless, the predicted collision outcome and the main morphological evolution remain consistent with the experiments. These comparisons show that the present framework can correctly reproduce bouncing at low Weber number and coalescence after crossing the regime boundary, without directly resolving the ultrathin gas film. This validates the effectiveness of the proposed method in handling the bouncing--coalescence mode transition.

\begin{figure}[!htbp]
    \centering
    \begin{minipage}[b]{0.48\columnwidth}
        \centering
        \includegraphics[width=\linewidth,keepaspectratio]{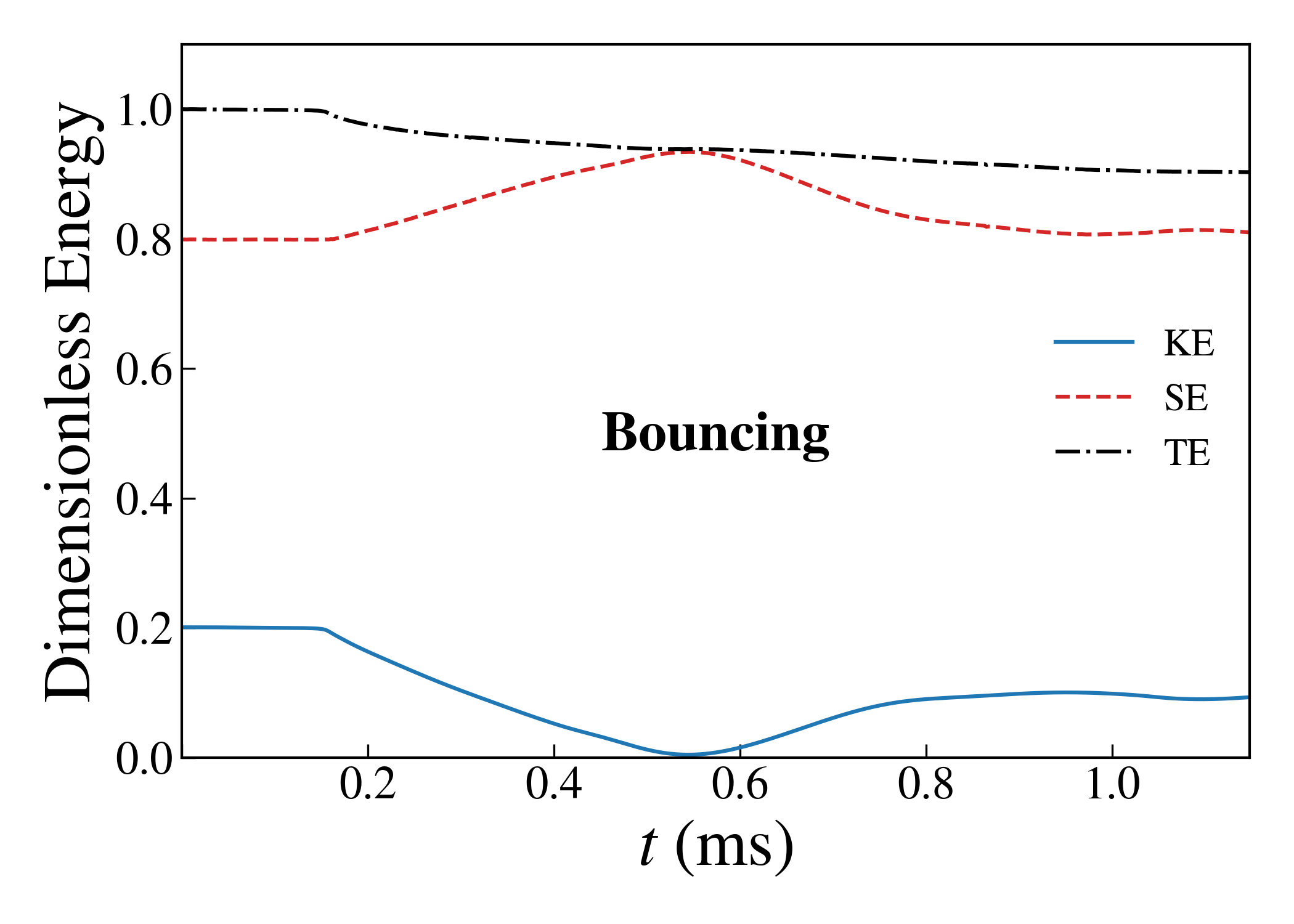}
        \centerline{(a) Bouncing}
    \end{minipage}
    \hfill
    \begin{minipage}[b]{0.48\columnwidth}
        \centering
        \includegraphics[width=\linewidth,keepaspectratio]{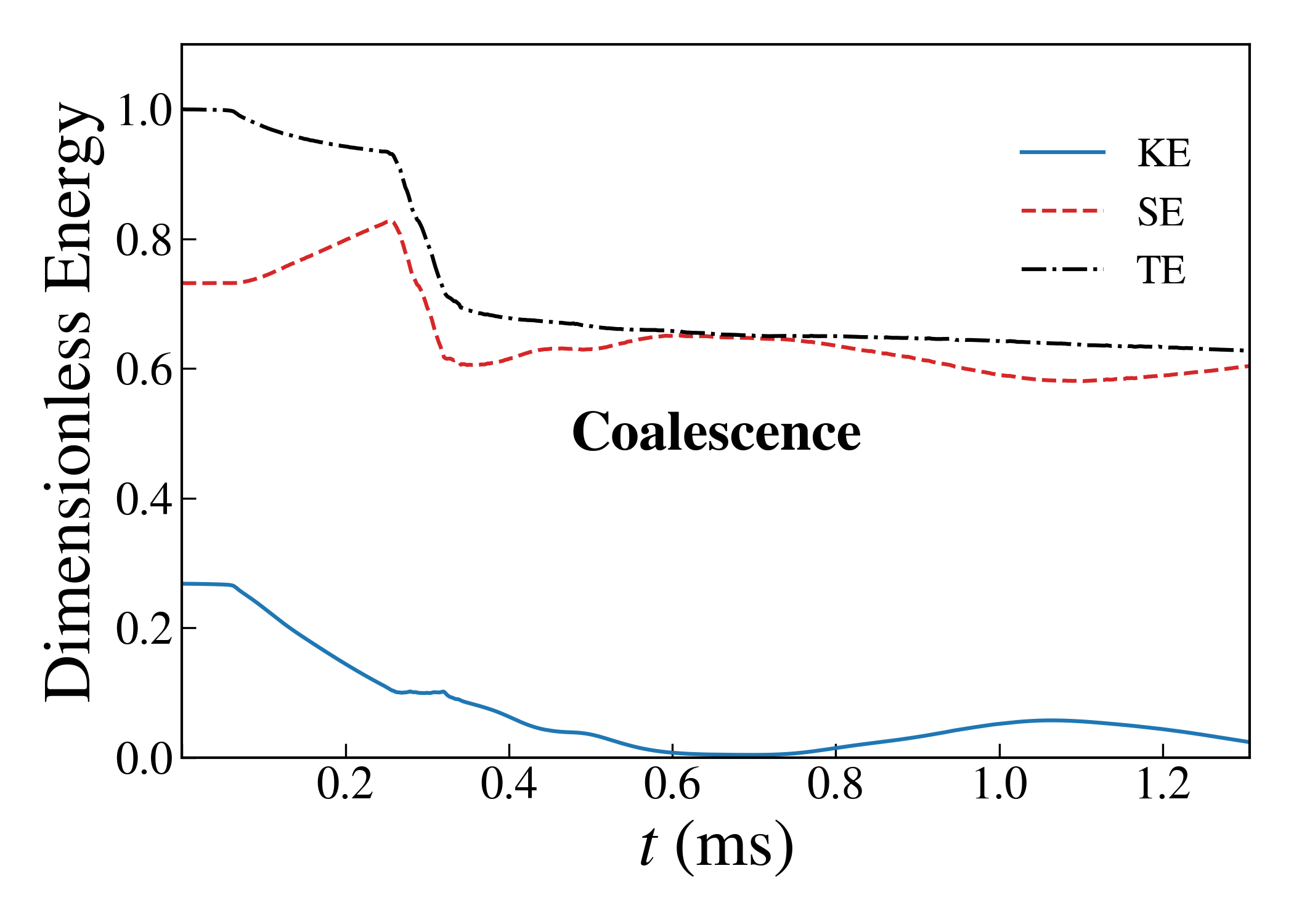}
        \centerline{(b) Coalescence}
    \end{minipage}
    \caption{Comparison of the temporal evolution of normalized energies during binary droplet collisions.
    (a) Normalized kinetic energy ($KE$), surface energy ($SE$), and total energy ($TE$) for the bouncing case at $We=9.33$;
    (b) corresponding energy evolution for the coalescence case at $We=13.63$.
    The bouncing case shows continuous and smooth energy conversion, whereas the stepwise changes in surface energy and total energy in the coalescence case correspond to the criterion-triggered topological merging operation.}
    \label{fig:energy_evolution}
\end{figure}

To further elucidate the difference between the two collision modes from an energetic perspective and to examine the reasonableness of the numerical model in representing energy conversion and dissipation, the dimensionless energy evolution during collision is analyzed. Figure~\ref{fig:energy_evolution} shows the temporal variations of the normalized kinetic energy $KE$, surface energy $SE$, and total energy $TE=KE+SE$ of the liquid phase for the bouncing case ($We=9.33$) and the coalescence case ($We=13.63$).

For the low-Weber-number bouncing case, the energy conversion process is continuous and smooth. During the initial approach stage, both the kinetic and surface energies vary only slightly. As the droplets compress each other, the kinetic energy is rapidly converted into surface energy, which drives pronounced radial spreading. At approximately $t\approx0.55~{\rm ms}$, the system reaches its maximum deformation, where the kinetic energy attains a minimum and the surface energy reaches a maximum. This moment corresponds to the maximum spreading state shown in Figs.~\ref{fig:pan_morphology}(a) and \ref{fig:pan_morphology}(b). Because the present algorithm maintains the independence of the multiple-VOF tracer fields in this case, no topological merging occurs. The surface energy stored in the deformed droplets is then smoothly converted back into kinetic energy under the action of surface tension, driving recoil and final bouncing. Throughout the process, the total energy decreases only gradually and continuously because of viscous dissipation within the liquid and in the gas-film region.

When inertia becomes sufficiently strong and the system enters the coalescence regime ($We=13.63$), the energy evolution exhibits distinctly different behavior. During the early stage of collision, the conversion from kinetic energy to surface energy is broadly similar to that in the bouncing case. However, at approximately $t\approx0.25~{\rm ms}$, the energy curves display a pronounced discontinuity. This moment corresponds to the topological merging operation triggered by the algorithm, i.e., the establishment of a connected liquid bridge through $c_{\rm new}\leftarrow c_1+c_2$. Physically, the connection of two previously independent interfaces leads to a discrete reduction in the total interfacial area, and therefore the surface energy $SE$ and total energy $TE$ in Fig.~\ref{fig:energy_evolution}(b) undergo a stepwise decrease.

It should be emphasized that the stepwise variations of surface energy and total energy in the coalescence case mainly reflect the unresolved bridge-establishment process represented by the topological switching operation. Since the present model does not directly resolve nanoscale gas-film rupture or the early coalescence dynamics in which the bridge radius grows from molecular to continuum scales, the abrupt energy drop should not be interpreted as a direct numerical resolution of early bridge growth. Instead, it should be regarded as a numerical diagnostic associated with the change in interfacial connectivity under the prescribed topological criterion. In other words, Fig.~\ref{fig:energy_evolution}(b) shows that the model has switched from two independent interfaces to one connected interface once the coalescence criterion is satisfied, rather than having resolved the full microscopic dynamics of early bridge growth.

After coalescence is triggered, the merged droplet enters a large-deformation nonlinear oscillatory stage dominated by surface tension. Because stronger internal recirculation and shear develop inside the merged droplet, the subsequent viscous dissipation is more pronounced, and the total energy continues to decrease. Overall, Fig.~\ref{fig:energy_evolution} shows that the present model can reasonably represent the conversion among kinetic energy, surface energy, and total energy under different collision modes. At the same time, the energy discontinuity in the coalescence case clearly illustrates the physical role of the present method, namely, reconstructing the macroscopic coalescence process through criterion-driven topological switching rather than directly resolving the microscopic details of gas-film rupture and early bridge growth.

\subsection{Validation of the machine-learning regime map and topological criterion}
\label{subsec:ml_phase_map_validation}

\begin{figure*}[!t]
    \centering
    \includegraphics[width=0.90\textwidth,keepaspectratio]{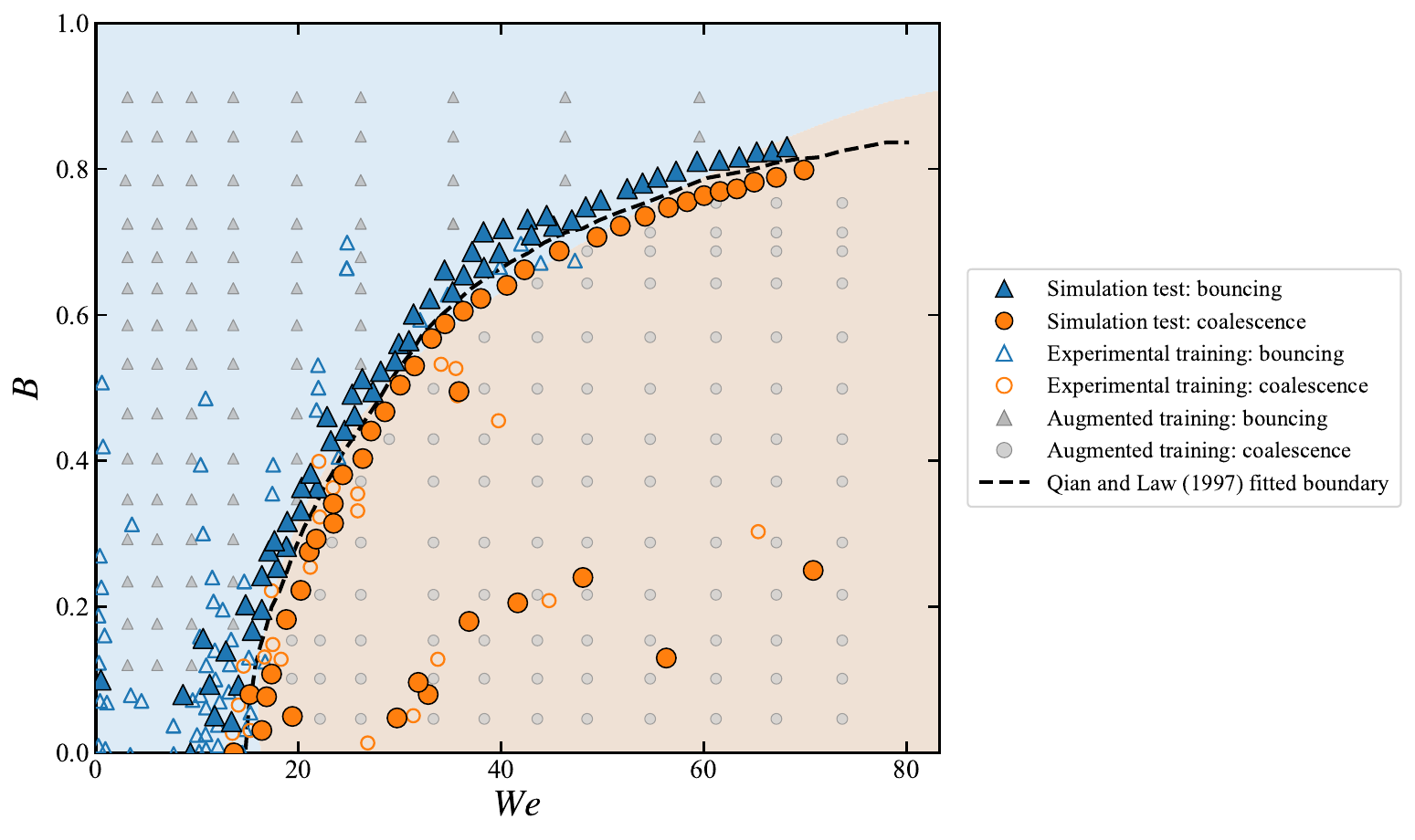}
   \caption{Machine-learning-predicted $We$--$B$ regime map for binary droplet collisions.
The light-blue and light-orange backgrounds denote the predicted bouncing and coalescence regions, respectively, and their interface corresponds to the classifier decision boundary.
The black dashed line denotes the experimental fitting boundary of Qian and Law~\cite{Qian1997}.
Open symbols denote experimental training data from Qian and Law~\cite{Qian1997} and Huang and Pan~\cite{Huang2021} ($66$ bouncing and $26$ coalescence samples), gray symbols denote extended training data generated from the experimental boundary curve ($88$ bouncing and $103$ coalescence samples), and filled symbols denote the independent simulation test set ($60$ bouncing and $45$ coalescence cases).}
    \label{fig:ml_phase_map}
\end{figure*}

To examine the feasibility of the machine-learning-assisted topological criterion for binary droplet collisions, a $We$--$B$ regime-map validation is constructed using experimental data, boundary-extended data, and an independent numerical simulation test set. The real experimental training data are taken from bouncing and coalescence cases reported by Qian and Law~\cite{Qian1997} and Huang and Pan~\cite{Huang2021}. Because the experimental data are sparse in some regions of the $We$--$B$ plane, a limited number of extended samples are generated from the experimental boundary curve to constrain the decision boundary of the machine-learning model in sparsely sampled regions. These extended samples are used only as low-weight auxiliary training data and are not treated as independent test data. Additional Basilisk simulation cases are then used as the test set to examine whether the trained classifier can provide reasonable predictions for bouncing and coalescence cases not included in training.

Figure~\ref{fig:ml_phase_map} shows the $We$--$B$ regime map predicted by the machine-learning model. The light-blue and light-orange backgrounds denote the predicted bouncing and coalescence regions, respectively. The boundary between the two background colors corresponds to the classifier decision boundary. The black dashed line denotes the experimental fitting boundary proposed by Qian and Law~\cite{Qian1997}. Open blue triangles and open orange circles represent bouncing and coalescence data in the real experimental training set, respectively. Gray triangles and gray circles denote the extended bouncing and coalescence training samples generated from the experimental boundary curve. Filled blue triangles and filled orange circles denote bouncing and coalescence samples in the present simulation test set.

As shown in Fig.~\ref{fig:ml_phase_map}, the machine-learning decision boundary follows the overall trend of the experimental fitting boundary of Qian and Law~\cite{Qian1997}. Within the bouncing/coalescence binary classification considered here, regions with larger $B$ mainly correspond to bouncing, whereas regions with smaller $B$ and moderate $We$ are more prone to coalescence. Compared with a single empirical fitting boundary, the machine-learning model learns the distribution of bouncing and coalescence regions directly from discrete labeled data, and therefore can provide a nonlinear boundary that better follows the local distribution of samples in data-rich regions. The gray extended samples help smooth and regularize the decision boundary. Because these samples are assigned lower training weights, the model boundary remains mainly controlled by the real experimental data.

More importantly, the simulation test set shown in Fig.~\ref{fig:ml_phase_map} is not involved in classifier training. The bouncing cases in the test set mainly fall within the machine-learning-predicted bouncing region, whereas the coalescence cases mainly fall within the predicted coalescence region. This indicates that the present machine-learning criterion based on $We$ and $B$ can provide reasonable classification for independent numerical cases and can serve as an executable criterion in the multiple-VOF topological-control framework.

In the numerical implementation, the machine-learning regime map in Fig.~\ref{fig:ml_phase_map} is not used only for offline classification. Instead, it is embedded into the topological-decision process after interfacial approach. When two droplet interfaces enter the near-contact region, the program extracts the current $We$ and $B$, and the machine-learning classifier returns either a bouncing or coalescence label. If the classifier returns a bouncing label, the two VOF tracer fields remain separated and continue to evolve independently. If a coalescence label is returned, the program triggers the topological merging operation $c_{\rm new}=\min(c_1+c_2,1)$. In this way, the prediction of the machine-learning model is directly converted into a topological-control instruction in the VOF solver.

The purpose of Fig.~\ref{fig:ml_phase_map} is to validate whether the machine-learning regime map can serve as an executable criterion in the multiple-VOF topological-control framework. The Weber number $We$ and impact parameter $B$ are selected as input features because they are the most commonly used control parameters in binary droplet collision regime maps and can be directly related to the bouncing and coalescence labels in experimental data. The results show that the classifier provides clear topological-decision regions in the $We$--$B$ plane and converts the physical partition in the experimental regime map into interfacial-control instructions for the numerical solver. This case therefore demonstrates the effective coupling between the data-driven criterion and the multiple-VOF topological-control framework.

\section{Extended application of the model: droplet impact on a liquid pool}
\label{sec:droplet_pool_extension}

The previous section validated the applicability of the criterion-driven multiple-VOF topological-control framework to binary droplet collisions. In this section, the framework is further extended to droplet impact on a liquid pool. Compared with binary droplet collisions, droplet--pool interaction involves stronger coupling with the deformable free surface. The initial kinetic energy of the impacting droplet is not only converted into deformation energy and viscous dissipation within the droplet, but also drives pool-surface depression, bouncing, and capillary-wave propagation. Therefore, whether the droplet eventually bounces or coalesces is not determined solely by the initial impact inertia. It is also closely related to thin-gas-film drainage between the droplet and the pool, the duration of near contact, and subsequent re-approach events.

Based on this physical distinction, a two-step validation strategy is adopted. First, a well-documented low-Weber-number non-coalescent bouncing case from the literature is selected as a benchmark to examine the ability of the present VOF solver to capture droplet deformation, pool-surface response, and bouncing trajectory. This step does not involve calibration of the drainage-time criterion. Its purpose is to confirm that the baseline flow solver and the multiple-VOF non-coalescence treatment can stably reproduce droplet--pool coupled dynamics. Subsequently, the effective drainage-time criterion is introduced using experimental cases, and the complete sequence from first bouncing to re-approach and final coalescence after droplet impact on a liquid pool is further examined.

The benchmark case is a low-Weber-number bouncing case in which a deionized water droplet impacts normally on a deep pool of the same liquid. The impacting droplet has a radius of $R=0.35~{\rm mm}$ and an initial impact velocity of $U_0=0.3855~{\rm m\,s^{-1}}$. In this study, the dimensionless parameters are defined using the droplet radius as the characteristic length,
\begin{equation}
We=\frac{\rho_l U_0^2 R}{\sigma}, \qquad
Bo=\frac{\rho_l g R^2}{\sigma}, \qquad
Oh=\frac{\mu_l}{\sqrt{\rho_l \sigma R}},
\label{eq:pool_dimensionless_numbers}
\end{equation}
where $\rho_l$, $\mu_l$, and $\sigma$ are the liquid density, dynamic viscosity, and surface tension coefficient, respectively. The corresponding parameters are $We=0.7$, $Bo=0.017$, and $Oh=0.006$, consistent with the low-Weber-number bouncing case of droplet impact on a deep liquid pool reported by Alventosa et al. \cite{Alventosa2023}. This case lies in an inertia--capillary-dominated regime, where gravity and viscous effects are relatively weak. Nevertheless, the coupled deformation between the droplet and the pool still strongly affects the bouncing trajectory and near-contact duration.

\begin{figure*}[!t]
    \centering
    \includegraphics[width=0.90\textwidth,keepaspectratio]{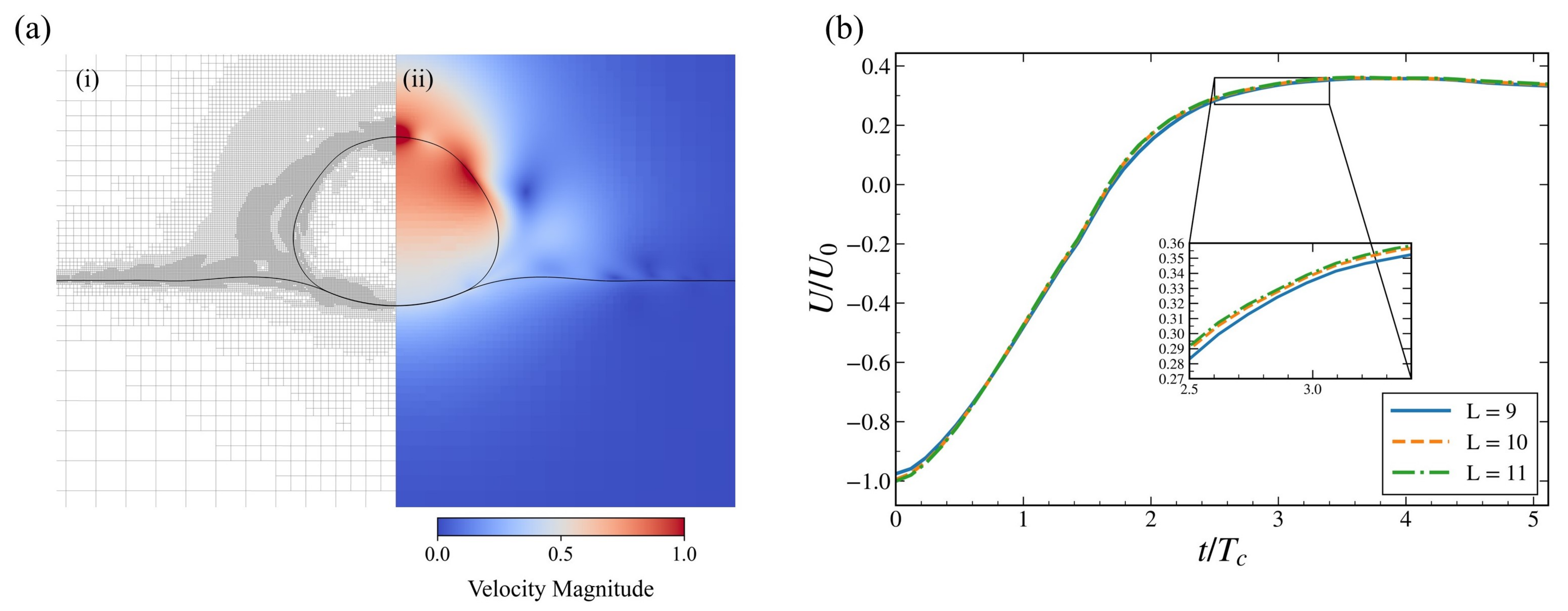}
  \caption{Local dynamics and grid-independence validation for the classical low-Weber-number bouncing case of droplet impact on a liquid pool.
  (a) Local flow structure at $t/T_c=1.00$, where (i) shows the adaptive mesh refinement (AMR) distribution and (ii) shows the velocity-magnitude contour;
  (b) temporal evolution of the normalized axial droplet velocity $U/U_0$ at different maximum mesh refinement levels, $L_{\max}=9$, $10$, and $11$.}
  \label{fig:pool_mesh_grid}
\end{figure*}

The simulations are performed in an axisymmetric $(r,z)$ plane. The computational domain has a size of $20R\times20R$, and the initial pool depth is set to $10R$, so as to reduce the influence of the bottom and lateral boundaries on the impact process. The droplet and the pool are represented by independent VOF tracer fields. In this benchmark bouncing case, the topological outcome is prescribed as non-coalescence. Therefore, when the interfaces approach each other, the droplet and pool tracer fields are kept independent to avoid nonphysical numerical coalescence caused by grid-scale contact. This treatment is consistent with the central idea of the proposed topological-control framework, namely, separating the solution of macroscopic interfacial motion from the topological decision of whether the interfaces are allowed to connect.

The initial mesh is uniformly divided into $2^8\times2^8$ cells, with pre-refinement applied near the droplet contour, the pool free surface, and the near-contact region between the two interfaces. The minimum adaptive refinement level is set to $L_{\min}=6$. To examine the effect of spatial resolution, the maximum refinement level is varied as $L_{\max}=9$, 10, and 11. Adaptive mesh refinement is performed using wavelet-based error estimates of the volume-fraction fields and the velocity field. The error thresholds are $10^{-5}$ for the volume-fraction fields and $10^{-3}$ for the velocity components. The time step is determined by both the CFL condition and the capillary stability constraint.

For quantitative comparison with the literature results, time is nondimensionalized by the radius-based capillary time,
\begin{equation}
T_c=\sqrt{\frac{\rho_l R^3}{\sigma}}.
\label{eq:radius_capillary_time_pool}
\end{equation}
The top and bottom positions of the droplet are defined as the highest and lowest axial positions of the droplet interface, respectively. The droplet centroid is obtained by integrating the volume-fraction field over the connected droplet region. The pool free-surface height is taken as the pool-interface position on the axis. The maximum droplet width $w$ is defined as the maximum radial width of the droplet interface at a given time. Accordingly, the trajectory and width are reported as $z/R$ and $w/R$, respectively.

Figure~\ref{fig:pool_mesh_grid}(a) shows the local flow structure at $t/T_c=1.00$. Figure~\ref{fig:pool_mesh_grid}(a)(i) shows the adaptive mesh distribution, and Fig.~\ref{fig:pool_mesh_grid}(a)(ii) shows the velocity-magnitude contour. The high-resolution cells are mainly concentrated near the droplet bottom, the depressed pool free surface, and the near-contact region between the two interfaces. These regions correspond to large interfacial curvature, strong velocity gradients, and pronounced pressure response, and therefore represent the main region of droplet--pool coupled dynamics. The velocity field further shows that, after the impacting droplet compresses the pool free surface, the local flow expands radially outward and gradually transitions to surface-tension-driven recoil and bouncing in the subsequent stage.

Figure~\ref{fig:pool_mesh_grid}(b) compares the temporal evolution of the normalized axial droplet velocity $U/U_0$ at different maximum refinement levels. The three results are consistent during the impact, deceleration, and bouncing stages, with only small differences near local extrema. When $L_{\max}$ is increased from 9 to 10, the velocity curve already shows clear convergence. Further increasing the refinement level to $L_{\max}=11$ produces only limited changes. Therefore, $L_{\max}=10$ is used as the main computational setting in the subsequent comparisons of morphology, trajectory, and maximum width for the benchmark case, so as to balance numerical accuracy and computational cost.

\subsection{Validation of morphological evolution for a classical low-Weber-number bouncing case}
\label{subsec:pool_benchmark_shape}

\begin{figure*}[!t]
    \centering
    \includegraphics[width=0.95\textwidth,keepaspectratio]{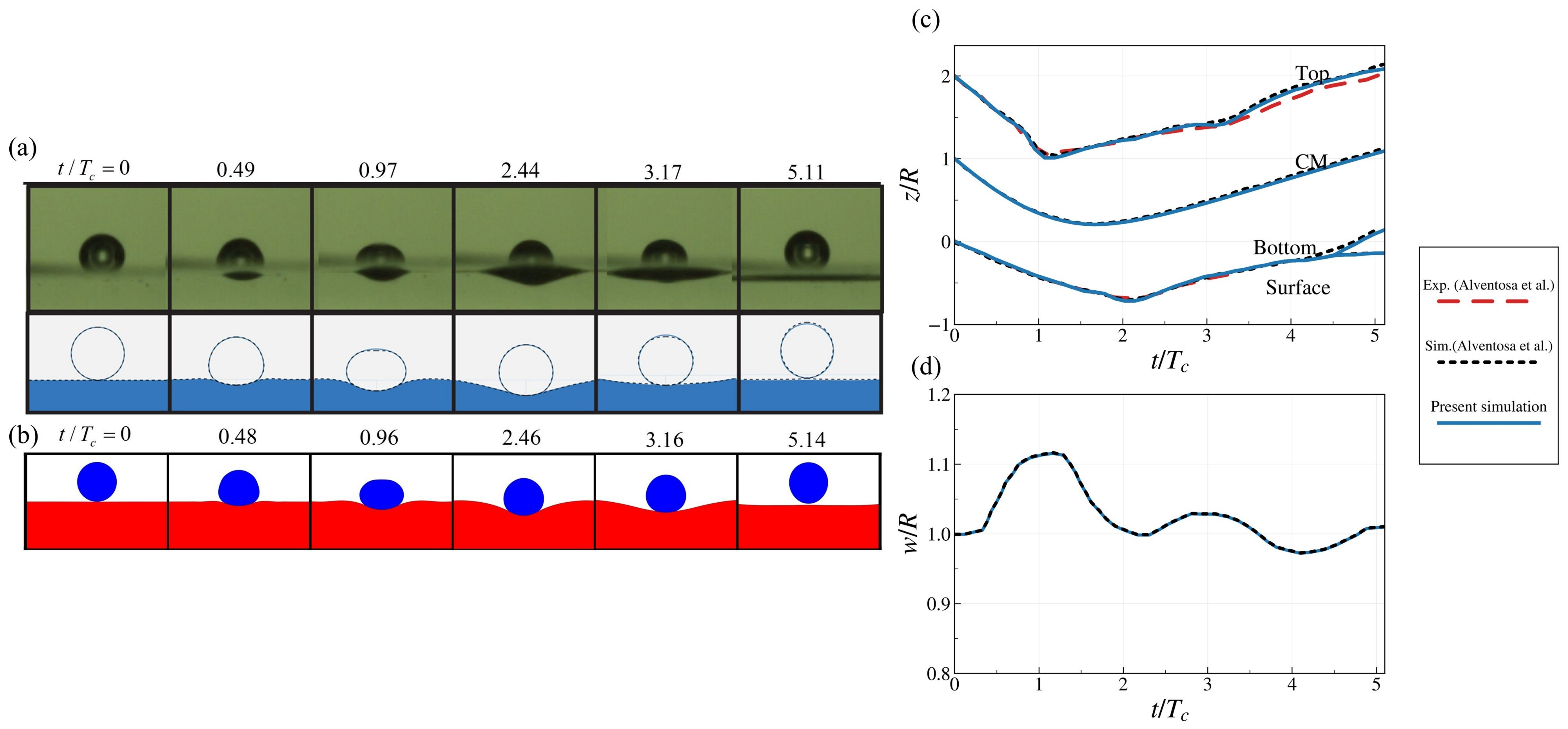}
    \caption{Benchmark validation for a classical low-Weber-number bouncing case of droplet impact on a liquid pool, with $We=0.7$, $Bo=0.017$, and $Oh=0.006$.
    (a) Morphological evolution reported by Alventosa et al.~\cite{Alventosa2023};
    (b) present numerical results;
    (c) comparison of the dimensionless trajectories of the droplet top, droplet bottom, pool free surface, and droplet center of mass;
    (d) comparison of the temporal evolution of the maximum droplet width $w/R$.}
    \label{fig:alventosa_validation}
\end{figure*}

After completing the numerical setup and grid-independence examination, the present solver is further applied to validate the morphological evolution of a classical low-Weber-number bouncing case of droplet impact on a liquid pool. Figures~\ref{fig:alventosa_validation}(a) and \ref{fig:alventosa_validation}(b) show the interfacial evolution sequences reported by Alventosa et al.~\cite{Alventosa2023} and obtained from the present simulation, respectively. The dimensionless parameters of this case are $We=0.7$, $Bo=0.017$, and $Oh=0.006$.

At the early impact stage, the droplet first compresses the pool free surface and produces a local depression. The droplet bottom then spreads and deforms together with the pool surface. As the contact region recoils, the droplet begins to detach from the pool surface and enters the bouncing stage. At later times, the droplet gradually recovers toward a nearly spherical shape, while the pool surface retains finite-amplitude wave motion. As shown in Figs.~\ref{fig:alventosa_validation}(a) and \ref{fig:alventosa_validation}(b), the present simulation reproduces the main morphological stages of this classical case. The droplet enters a clear depression stage near $t/T_c\approx1$, reaches a pronounced pool-surface depression and droplet deformation near $t/T_c\approx2.5$, and then leaves the pool surface during bouncing. Both the relative motion between the droplet bottom and the pool surface and the recovery of the droplet shape after bouncing agree well with the literature results.

For droplet impact on a liquid pool, agreement in the morphological sequence has direct physical significance. Rebound in this problem is not controlled solely by the deformation of the droplet itself, but is jointly determined by the droplet, the pool free surface, and the local flow between them. Therefore, the correspondence between Figs.~\ref{fig:alventosa_validation}(a) and \ref{fig:alventosa_validation}(b) indicates not only that the interface capturing is stable, but also that the amplitude and timing of the pool-surface response are reasonably represented. This validation provides the necessary numerical basis for the subsequent criterion-based topological-control modeling.

To further examine the predictive capability of the present numerical framework for this benchmark case, Fig.~\ref{fig:alventosa_validation}(c) compares the dimensionless trajectories of the droplet top, droplet bottom, pool free surface, and droplet center of mass, while Fig.~\ref{fig:alventosa_validation}(d) shows the temporal evolution of the maximum droplet width $w/R$. Compared with morphological sequences alone, these quantities more directly reflect the dynamic responses of the droplet and pool during impact, depression, recoil, and bouncing.

As shown in Fig.~\ref{fig:alventosa_validation}(c), the overall trends of the four characteristic trajectories obtained in the present simulation are consistent with the experimental and numerical results of Alventosa et al.~\cite{Alventosa2023}. The droplet top position decreases rapidly after impact and then rises after reaching a local minimum. The positions of the droplet bottom and the pool free surface more directly reflect the depression and recovery of the near-contact region. The center-of-mass trajectory characterizes the recovery of the droplet momentum and the bouncing height. The present results capture both the main trends of these quantities and the timing of the key turning points, indicating that the solver is comparable to the literature results not only in local interfacial morphology but also in the global dynamic response.

The maximum-width result in Fig.~\ref{fig:alventosa_validation}(d) further validates the predicted deformation intensity of the droplet. The maximum width directly reflects the competition between inertial spreading and capillary recoil, and is sensitive to the resolution of local curvature and the accuracy of the velocity field. The present results agree well with the literature results in terms of the peak location, peak magnitude, and subsequent oscillation trend. This indicates that the current AMR strategy can not only predict the overall trajectory reasonably, but also maintain sufficient local spatial resolution during the stage of strong droplet deformation.

Combining the results in Figs.~\ref{fig:alventosa_validation}(c) and \ref{fig:alventosa_validation}(d), the present axisymmetric VOF framework can be considered capable of reproducing the main observable quantities in the classical low-Weber-number bouncing case of droplet impact on a liquid pool. Because this benchmark involves droplet-bottom depression, pool free-surface response, and droplet-shape recovery after bouncing, this validation provides important support for the subsequent analysis of droplet impact on a liquid pool.

\subsection{Experimental and numerical comparison of the complete impact process for water and silicone-oil droplets}
\label{subsec:pool_full_process}

After establishing the numerical baseline using the classical bouncing benchmark, we further examine whether the proximity criterion and the effective drainage-time criterion proposed in Sec.~2 can describe the complete topological evolution after droplet impact on a liquid pool. Unlike the preceding benchmark, which mainly tests the ability of the solver to predict non-coalescent bouncing dynamics, this section focuses on whether the algorithm can continuously describe the first impact, short-time bouncing, re-approach, and final coalescence within the same simulation. To assess the applicability of the method under different liquid-property conditions, two representative cases are considered: a deionized water droplet and a $5~{\rm cSt}$ silicone-oil droplet. The former has weaker viscous dissipation and weaker gravitational effects, whereas the latter has higher viscosity and a more pronounced free-surface deformation response. These two cases therefore test the stability of the proposed topological-control framework over different material-property scales.

\begin{figure}[htbp]
  \centering
  \includegraphics[width=8.5cm]{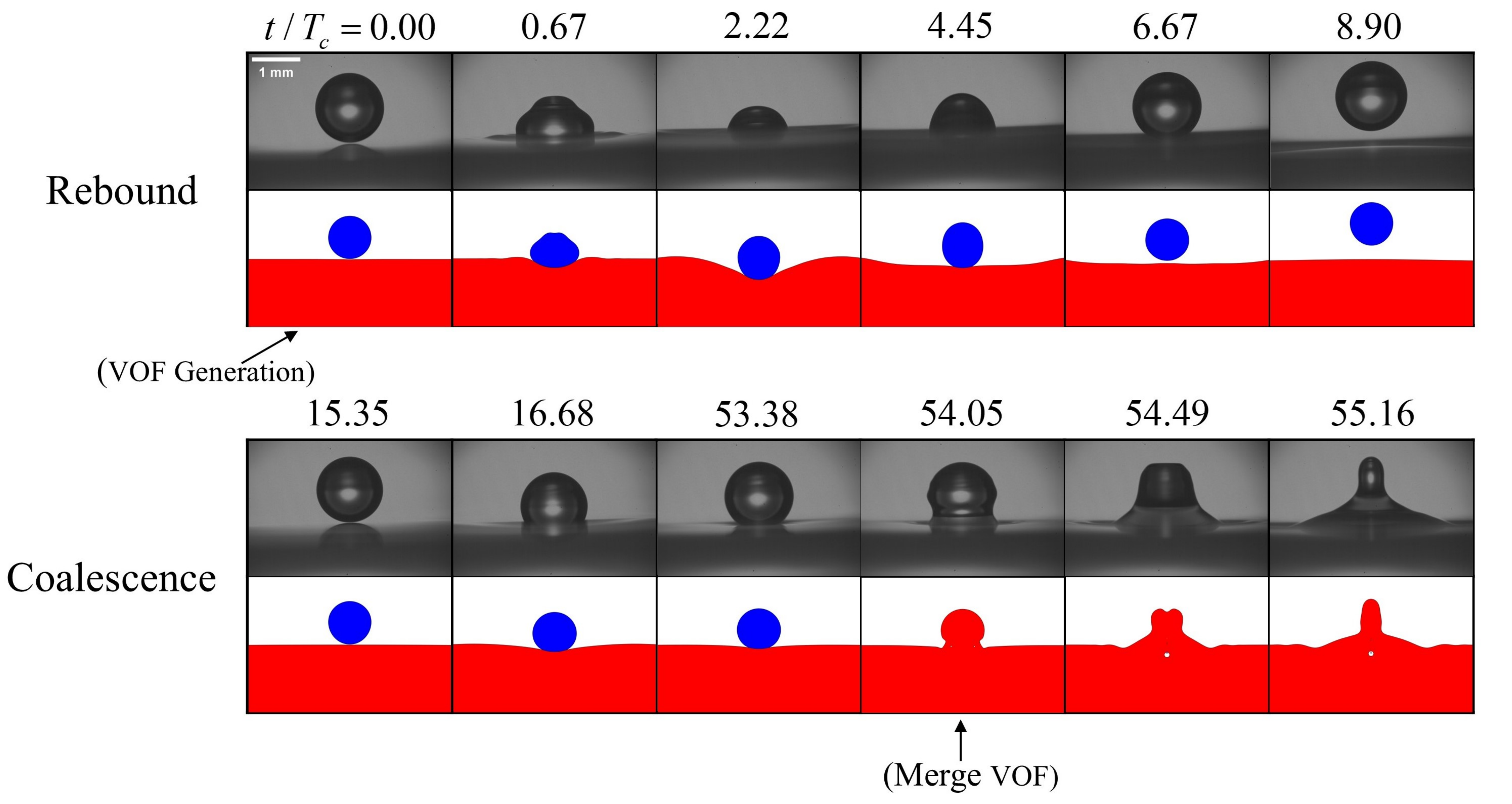}
  \caption{Experimental and numerical comparison of the complete impact process of a deionized water droplet on a liquid pool. The upper part shows the non-coalescent bouncing stage after the first impact, and the lower part shows the stage of re-approach and final coalescence after bouncing. In each group, the upper row shows the experimental images, and the lower row shows the numerical results. The droplet radius is $R=0.715~{\rm mm}$, and the initial impact velocity is $U_0=0.40~{\rm m\,s^{-1}}$. The dimensionless parameters are defined using the droplet radius and are $We=1.58$, $Oh=0.0043$, and $Bo=0.07$. The calibrated effective drainage-time threshold is $t_d=120.75~{\rm ms}$, corresponding to $C_d=53.7$.}
  \label{fig:water_sequence}
\end{figure}

Figure~\ref{fig:water_sequence} compares the experimental and numerical results for the complete impact process of a deionized water droplet on a liquid pool. In this case, the droplet radius is $R=0.715~{\rm mm}$, and the initial impact velocity is $U_0=0.40~{\rm m\,s^{-1}}$. Using the droplet radius as the characteristic length, the corresponding dimensionless parameters are $We=1.58$, $Oh=0.0043$, and $Bo=0.07$. Because $Oh$ is small, the impact process of the water droplet is mainly governed by inertia and surface tension, while viscous dissipation weakly suppresses the first bouncing stage. As a result, the droplet maintains a clear non-coalescent bouncing after its first approach to the pool.

As shown in Fig.~\ref{fig:water_sequence}, both the experiment and simulation exhibit a two-stage evolution after the water droplet impacts the pool. In the first stage, the droplet approaches the free surface and compresses it to form a local deformation near $t/T_c=0.67$. The droplet and pool then remain topologically separated and complete recoil and bouncing over the interval $t/T_c=2.22$ to $t/T_c=8.90$. In this stage, the numerical model keeps the droplet and pool tracer fields independent, so premature numerical coalescence does not occur. The experiment shows that the droplet recovers toward a nearly spherical shape after leaving the pool surface, and the numerical result reproduces the same morphological trend.

In the second stage, the bounced water droplet approaches the pool surface again under gravity. The experimental images show that the droplet re-enters the vicinity of the pool surface after $t/T_c=53.38$, forms a liquid bridge with the pool near $t/T_c=54.05$, and then enters the post-coalescence free-surface reconstruction stage over $t/T_c=54.49$ to $t/T_c=55.16$. The numerical results reproduce the same topological transition. Before re-approach, the droplet moves as an independent topological structure. Once the accumulated near-contact time reaches the effective drainage-time threshold, the algorithm triggers tracer-field merging, and the droplet and pool change from separated to connected states. This result shows that the present method can not only maintain non-coalescent bouncing after the first impact of the water droplet, but also trigger final coalescence during the subsequent re-contact stage according to the drainage-time criterion.

\begin{figure}[htbp]
  \centering
  \includegraphics[width=8.5cm]{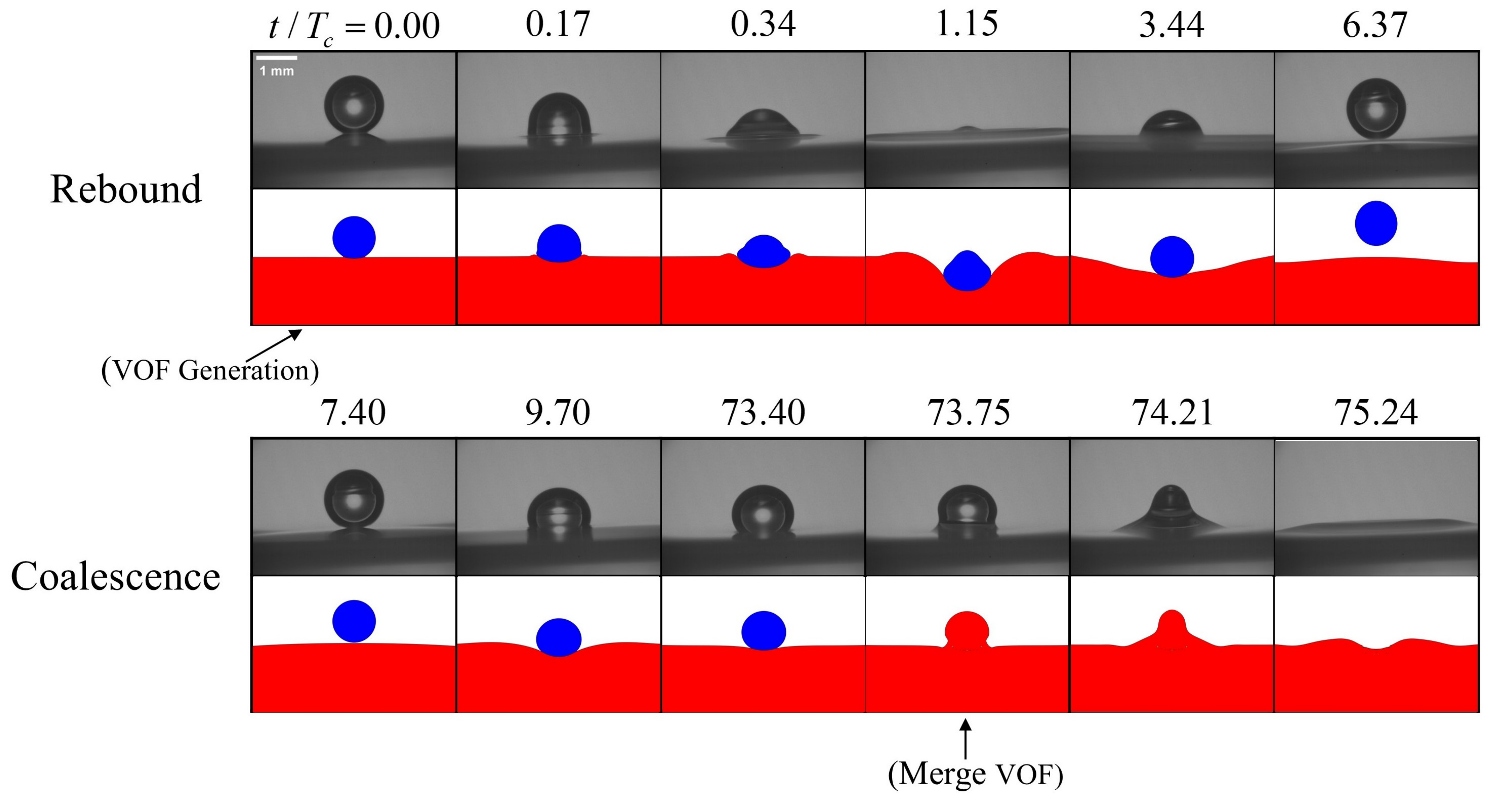}
  \caption{Experimental and numerical comparison of the complete impact process of a $5~{\rm cSt}$ silicone-oil droplet on a liquid pool. The figure shows the continuous evolution from the first impact, bouncing, and re-approach to final coalescence. The droplet radius is $R=0.74~{\rm mm}$, and the initial impact velocity is $U_0=0.49~{\rm m\,s^{-1}}$. The dimensionless parameters are defined using the droplet radius and are $We=8.32$, $Oh=0.0398$, and $Bo=0.252$. The calibrated effective drainage-time threshold is $t_d=319.75~{\rm ms}$, corresponding to $C_d=73.4$.}
  \label{fig:silicone_sequence}
\end{figure}

Figure~\ref{fig:silicone_sequence} presents the experimental and numerical comparison for the complete impact process of a $5~{\rm cSt}$ silicone-oil droplet on a liquid pool. In this case, the droplet radius is $R=0.74~{\rm mm}$, and the initial impact velocity is $U_0=0.49~{\rm m\,s^{-1}}$. Using the droplet radius as the characteristic length, the corresponding dimensionless parameters are $We=8.32$, $Oh=0.0398$, and $Bo=0.252$. Compared with the deionized-water case, the silicone-oil case has larger $Oh$ and $Bo$. Therefore, droplet recoil and pool-surface recovery are more strongly affected by viscous dissipation and gravity. In this case, the droplet also does not coalesce immediately after its first contact with the pool. Instead, it undergoes a continuous sequence of pool-surface depression, droplet recoil, short-time bouncing, re-approach to the surface, and final coalescence.

As shown in Fig.~\ref{fig:silicone_sequence}, the numerical results reproduce the key interfacial morphologies observed in the experiment. During the first impact stage, the droplet compresses the pool free surface downward under inertia, while the droplet itself undergoes clear deformation. Subsequently, the droplet and pool interfaces recoil under surface tension, and the droplet detaches from the pool surface and enters a short-time bouncing stage. During the later motion, the droplet approaches the pool surface again and establishes a topological connection with the pool after the effective drainage-time criterion is satisfied, leading to final coalescence. Compared with the water case, the silicone-oil droplet recovers more gradually after bouncing, and the pool-surface disturbance decays more evidently, consistent with its stronger viscous dissipation.

Combining Figs.~\ref{fig:water_sequence} and \ref{fig:silicone_sequence}, the present method can reproduce the complete topological evolution after droplet impact on a liquid pool under two different liquid-property conditions. The deionized-water case has lower $Oh$ and $Bo$, and the droplet maintains more pronounced inertia--capillary oscillations after the first bouncing. The silicone-oil case has higher $Oh$ and $Bo$, and the deformation and recovery of the droplet and pool free surface are more strongly modulated by viscous dissipation and gravity. Although the material properties, bouncing amplitudes, and re-contact times differ between the two liquids, the numerical results correctly reproduce the stage sequence of first impact, bouncing, re-approach, and final coalescence.

This result further indicates that the role of the present model is not limited to reproducing a specific silicone-oil case. Instead, by combining the proximity criterion with the effective drainage-time criterion, delayed coalescence during droplet impact on a liquid pool is reformulated as an executable topological-control problem. Compared with previous studies that mainly focused on a single bouncing event or the bouncing--coalescence transition condition, the present work emphasizes the reproduction of the complete temporal sequence within the same simulation. In other words, the model must not only maintain the topological separation between the droplet and the pool during the first near-contact event, but also actively trigger coalescence when the droplet approaches again and the drainage-time condition is satisfied. The results for the water and silicone-oil cases jointly show that the criterion-driven topological-control framework proposed in Sec.~2 can stably describe bouncing, re-contact, and final coalescence after droplet impact on a liquid pool under different liquid-property conditions.

\subsection{Temporal evolution of the droplet-top position}
\label{subsec:silicone_top_position}

\begin{figure}[!htbp]
  \centering
  \includegraphics[width=8cm]{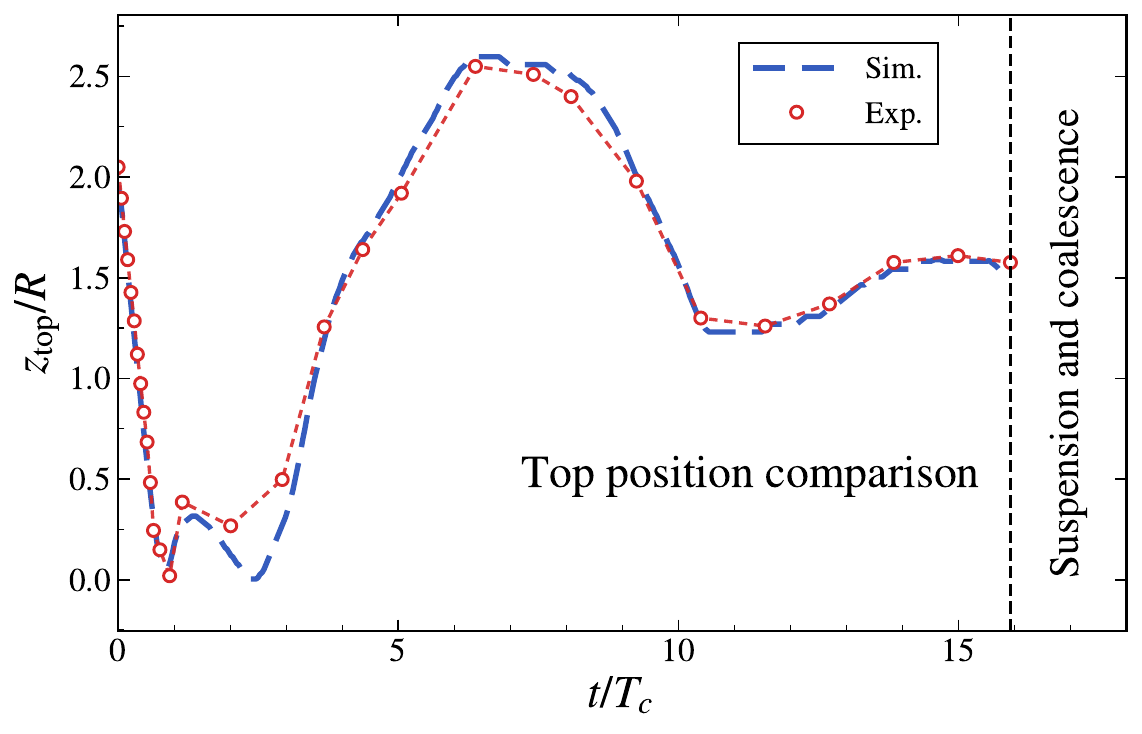}
  \caption{Temporal evolution of the droplet-top position relative to the initial pool surface during silicone-oil droplet impact on a liquid pool. Red open circles denote the experimental results, and the blue dashed line denotes the numerical results. The black dashed line marks the beginning of the small-amplitude oscillation stage near the free surface after the first bouncing; final coalescence occurs at a later time, as shown in Fig.~\ref{fig:silicone_sequence}.}
  \label{fig:silicone_top}
\end{figure}

To further quantify the comparison between the experiment and simulation, Fig.~\ref{fig:silicone_top} shows the temporal evolution of the top position of the silicone-oil droplet relative to the initial pool surface. During the sinking and near-contact stages, it is difficult to clearly identify the droplet center of mass and the lower droplet interface from the experimental images. Therefore, the droplet-top position is selected as the primary quantity for comparison between the experiment and simulation. This quantity can be identified more reliably throughout the evolution and provides a stable measure of the droplet trajectory during impact, bouncing, re-approach, and before final coalescence.

As shown in Fig.~\ref{fig:silicone_top}, the evolution of the droplet-top position can be divided into four stages. The first stage is the inertial depression stage, during which the droplet continues moving downward after impact and the top position decreases rapidly. The second stage is the capillary recovery stage. As the deformation energy stored in the droplet and pool free surface is released, the droplet detaches from the pool surface and rebounds upward, leading to an increase in the top position. The third stage is the re-approach stage, in which the droplet falls again under gravity and re-enters the near-contact region close to the pool surface. The fourth stage is the final coalescence stage. At this stage, the droplet establishes a topological connection with the pool, and the trajectory of the initially independent droplet terminates. The droplet-top curve is therefore ended at the coalescence time.

The experimental and numerical results show good agreement in the overall trend. The simulation reproduces the rapid downward motion after the first impact and captures the subsequent downward motion after bouncing. More importantly, the numerical result describes not only the interfacial dynamic response after the first impact, but also the complete trajectory before final coalescence. This indicates that the algorithm proposed in Sec.~2 is not limited to describing the short-time bouncing after the first impact. Instead, it can continuously track the near-contact state between the droplet and the pool throughout the computation and trigger a topological update when the effective drainage-time condition is satisfied.

After $t/T_c\approx16$, the droplet remains topologically separated from the pool and undergoes small-amplitude oscillations near the free surface. During this stage, the variation of $z_{\mathrm{top}}/R$ becomes weak and provides limited additional information about the main dynamics of the first impact and bouncing. Final coalescence occurs at a much later time, approximately $t/T_c\approx73.4$, as shown in Fig.~\ref{fig:silicone_sequence}. Therefore, Fig.~\ref{fig:silicone_top} mainly focuses on the impact, bouncing, and early relaxation stages, where the droplet-top position exhibits more pronounced changes.

It should be noted that, after coalescence occurs, the droplet-top position is no longer used to describe the subsequent interfacial evolution. Once the droplet and pool are topologically connected, the droplet no longer exists as an independent object. The subsequent motion corresponds to the evolution of the combined free surface rather than the trajectory of an individual droplet. Therefore, terminating the droplet-top curve at the coalescence time is more consistent with the definition of this quantity and helps clearly identify the instant of topological transition.

\begin{figure*}[!t]
    \centering
    \includegraphics[width=0.90\textwidth,keepaspectratio]{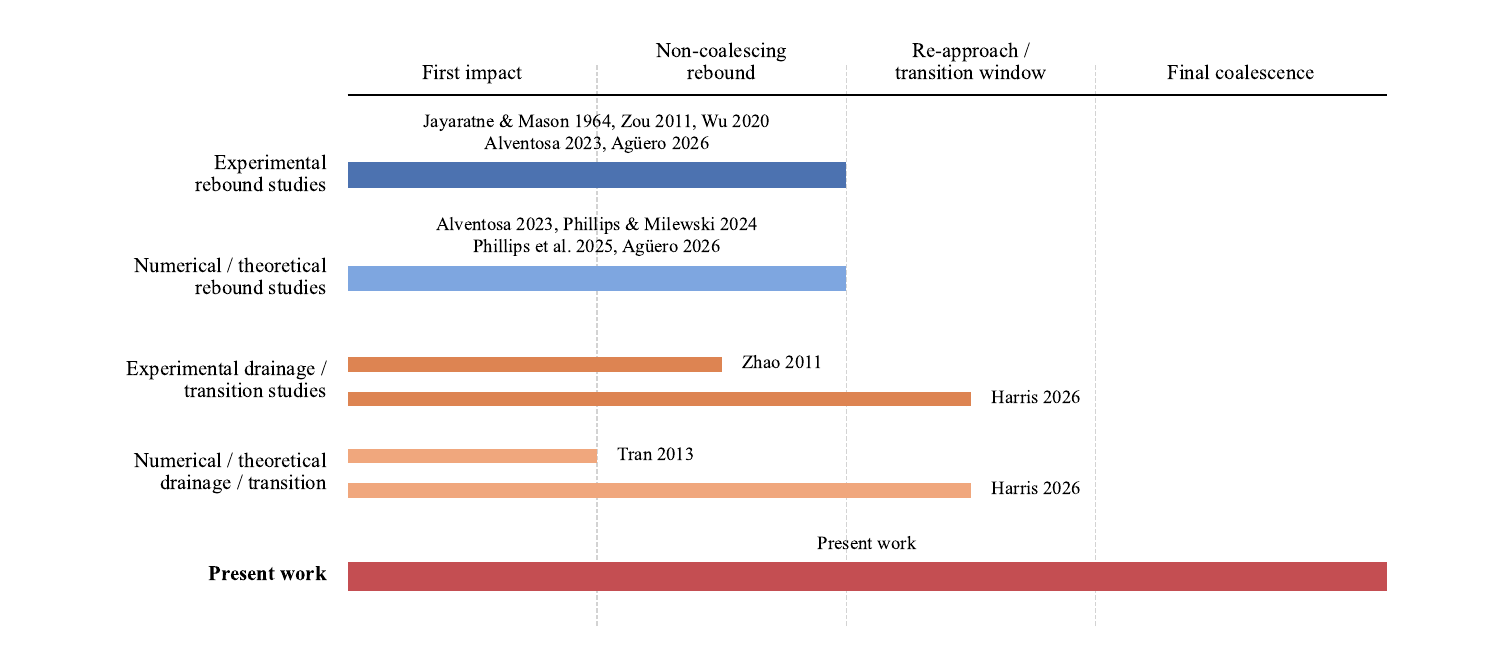}
    \caption{Schematic process coverage of representative studies on droplet impact on liquid pools~\cite{Jayaratne1964,Zou2011,Zhao2011,Wu2020,Tran2013,Alventosa2023,Aguero2026LowWeber,Harris2026MovingPools,Phillips2024Lubrication,Phillips2025TwoDimensional}.
    The horizontal direction denotes successive stages, including first impact, non-coalescent bouncing, re-approach or transition window, and final coalescence.
    Blue bars denote studies mainly focused on non-coalescent bouncing dynamics, with experimental and numerical/theoretical works distinguished.
    Orange bars denote studies mainly focused on gas-film drainage, film rupture, or the bouncing--coalescence transition, again distinguishing experimental and numerical/theoretical works.
    The red bar denotes the present work. The diagram is intended for a qualitative comparison of the process stages covered by different studies and does not represent a strict quantitative time scale.}
    \label{fig:process_coverage}
\end{figure*}

\subsection{First impact, short-time bouncing, re-approach, and final coalescence}
\label{subsec:process_level_comparison}

To further clarify the relation between the present work and previous studies on droplet impact on liquid pools, Fig.~\ref{fig:process_coverage} schematically summarizes the process stages covered by representative studies. Existing works can be broadly divided into two categories. The first category mainly focuses on non-coalescent bouncing dynamics after droplet impact, including the early experiments of Jayaratne and Mason~\cite{Jayaratne1964}, the bouncing experiments of Zou et al.~\cite{Zou2011} and Wu et al.~\cite{Wu2020}, and the experimental, theoretical, and numerical studies of low-Weber-number bouncing on deep liquid pools by Alventosa et al.~\cite{Alventosa2023} and Agüero et al.~\cite{Aguero2026LowWeber}. The second category focuses more on entrapped gas-film drainage, film rupture, and the bouncing--coalescence transition, including the experimental study of bouncing and coalescence transition on deep liquid pools by Zhao et al.~\cite{Zhao2011}, the analysis of gas-film formation and rupture by Tran et al.~\cite{Tran2013}, and the experimental and direct numerical simulation study of the bouncing--coalescence transition on moving pools by Harris et al.~\cite{Harris2026MovingPools}. Compared with these studies, the present work further focuses on the complete evolution pathway in which a droplet undergoes first impact, short-time bouncing, re-approach, and final coalescence within the same simulation.

As shown in Fig.~\ref{fig:process_coverage}, the benchmark validation discussed above mainly corresponds to the first-impact and non-coalescent bouncing stages. It is used to establish the reliability of the present numerical framework in capturing droplet--pool coupled deformation, free-surface depression, and bouncing dynamics. The water and silicone-oil cases considered subsequently extend the analysis to the re-approach and final-coalescence stages. In other words, previous studies have mostly focused either on the bouncing stage itself or on the conditions and mechanisms of the bouncing-to-coalescence transition. Building on these works, the present study combines the proximity criterion and the effective drainage-time criterion developed in Sec.~2 to further examine the complete dynamic sequence of first impact, bouncing, re-approach, and final coalescence after droplet impact on a liquid pool. This comparison indicates that the main role of the present method is not to replace microscopic models of film drainage, but to provide a criterion-driven implementation for macroscopic VOF solvers that can cover the complete topological evolution pathway.

\section{Conclusions}
\label{sec:conclusions}

The bouncing and coalescence of colliding droplets are complex physical processes, and they involve the difficulty of capturing dynamic changes in interfacial topology. Methods such as mesh refinement, gas-film models, van der Waals molecular-force models, or empirical models cannot provide a general and effective way to solve the formation of complex collision regimes and the associated probabilistic behavior. In this work, we argue that, instead of expecting unlimited improvement of numerical algorithms to adapt to the accurate simulation of bouncing and coalescence, it is more appropriate to separate the numerical evolution of the interface from the physical decision-making process that determines whether bouncing or coalescence occurs. In this way, a more general simulation of bouncing and coalescence can be achieved within a finite computational cost.

In this study, the fusion and generation of VOF model are constructed to represent droplet coalescence and bouncing, respectively. The decision of whether bouncing or coalescence occurs is assigned to a machine-learning network that can learn and evolve. By fully utilizing the data-learning and generalization capability of the machine-learning network, the proposed learning and simulation framework can, in principle, be applied to the simulation of interfacial bouncing and coalescence phenomena. This framework also extends the capability of the VOF method in capturing interfacial problems. The free generation and annihilation of interfaces are beneficial for simulating more realistic interfacial processes, while avoiding the trap of pursuing infinitely high grid resolution. Meanwhile, because the network has an intrinsic probabilistic nature through stochastic connection and probability-based decision-making, it has a natural adaptability to the randomness of bouncing and coalescence phenomena.

Based on Basilisk, this work constructs an interfacial computation--learning framework. In the simulations of binary droplet collisions, the obtained results can well reflect bouncing and coalescence phenomena and provide a clear phase-map boundary. The results are consistent with previous experimental and numerical results. We further extend the simulation framework to the problem of droplet impact on a liquid pool, and the numerical results agree well with our experimental observations. At the same time, using the present simulation framework, we realize for the first time the complete numerical simulation of the droplet process of ``collision--rebound--re-collision--coalescence'' within the same simulation process, and the result is consistent with our experiment. Experimentally realizing the complete process of ``collision--bouncing--re-collision--coalescence'' within one process is also difficult; however, achieving the complete simulation of this process is even more challenging for previous numerical approaches. Therefore, there has been no precedent for such a simulation.

We believe that by fully utilizing the capability of machine learning in data mining and evolution of physical phenomena, and by appropriately transforming numerical algorithmic tools, it is possible to build an evolutionary simulation framework system in which different components play their respective strengths. Such a framework can adapt to the richness and diversity of physical phenomena. This work provides a useful example for the practice of this route.

\section*{Acknowledgements}
This work was supported by the National Natural Science Foundation of China-Yunnan Joint Fund Key Project, U2002214. The authors would like to express their sincere gratitude to Professors Jin-Song Zhang, Guo-Hui Hu, Zhe-Wei Zhou, and Qian Xu for valuable assistance.

\section*{DATA AVAILABILITY}
The data that support the findings of this study are available from the corresponding author upon reasonable request.

\bibliography{apssamp}

\end{document}